\documentclass[manuscript]{aastex63}
\usepackage{amsmath}
\usepackage{wasysym}
\received{December 11, 2020}
\revised{February 18, 2021}
\accepted{February 25, 2021}
\submitjournal{ApJ}

\shorttitle{Polytropic Filamentary Clouds}
\shortauthors{Kashiwagi \& Tomisaka}
\graphicspath{{./}{figures/}}
\begin{document}

\title{Magnetohydrostatic Equilibrium Structure and Mass of Polytropic Filamentary Cloud Threaded by Lateral Magnetic Field}

\correspondingauthor{Raiga Kashiwagi}
\email{raiga.kashiwagi@grad.nao.ac.jp}

\author[0000-0002-1461-3866]{Raiga Kashiwagi}
\affiliation{Department of Astronomical Science, School of Physical Sciences, The Graduate University for Advanced Studies (SOKENDAI), 2-21-1 Osawa, Mitaka, Tokyo 181-8588,
Japan}

\author[0000-0003-2726-0892]{Kohji Tomisaka}
\affiliation{Department of Astronomical Science, School of Physical Sciences, The Graduate University for Advanced Studies (SOKENDAI), 2-21-1 Osawa, Mitaka, Tokyo 181-8588,
Japan}
\affiliation{Division of Science, National Astronomical Observatory of Japan, 2-21-1 Osawa, Mitaka, Tokyo 181-8588, Japan}



\begin{abstract}
Filamentary structures are recognized as a fundamental component of interstellar molecular clouds in observations by the {\it Herschel} satellite. 
These filaments, especially massive filaments, often extend in a direction perpendicular to the interstellar magnetic field. Furthermore, the filaments sometimes have an apparently negative temperature gradient, that is, their temperature decreases towards the center.
In this paper, we study the magnetohydrostatic equilibrium state of negative-indexed polytropic gas with the magnetic field running perpendicular to the axis of the filament. 
The model is controlled by four parameters: center-to-surface density ratio ($\rho_c/\rho_s$), plasma $\beta$ of the surrounding gas, radius of the parent cloud $R'_0$ normalized by the scale height, and the polytropic index $N$. 
The steepness of the temperature gradient is represented by $N$. We found that the envelope of the column density profile becomes shallow when the temperature gradient is large.This reconciles the inconsistency between the observed profiles and those expected from the isothermal models.
We compared the maximum line-mass (mass per unit length), above which there is no equilibrium, with that of the isothermal non-magnetized filament.
We obtained an empirical formula to express the maximum line-mass of a magnetized polytropic filament as 
$\lambda_{\rm{}max}\simeq \left[{\left(\lambda_{0,\rm max}(N)/M_\odot~\rm{}pc^{-1}\right)^2+\left[5.9\left(1.0+1.2/N\right)^{1/2}\left({\Phi_{\rm{}cl}}/{1\rm{}\mu{}G\,pc}\right)\right]^2}\right]^{1/2}M_\odot~\rm{}pc^{-1}$, where $\lambda_{0,\rm{}max}(N)$ represents the maximum line-mass of the non-magnetized filament and $\Phi_{\rm{}cl}$ indicates one-half of the magnetic flux threading the filament per unit length.
Although the negative-indexed polytrope makes the maximum line-mass decrease compared with that of the isothermal model, a magnetic field threading the filament increases the line-mass.
\end{abstract}


\section{Introduction}\label{sec:intro}

Filamentary structures in molecular clouds are  recently attracting much attention among researchers aiming to understand the earliest phase of star formation.
The {\it Herschel} space observatory \citep{2010A&A...518L...1P} has revealed that the filamentary structure is a basic component of nearby molecular clouds by observing thermal dust emissions in the far infrared and submillimeter ranges \citep{2010A&A...518L.102A}. 
Not only active star-forming regions, such as Aquila \citep{2010A&A...518L.103M}, Taurus \citep{2013A&A...550A..38P}, and IC5146 \citep{2011A&A...529L...6A}, but also inactive ones such as Polaris \citep{2010A&A...518L..92W} have indicated the presence of a filament system.

The magnetic field structure in molecular clouds can be studied in several ways, such as the near-infrared polarization of background stars and the polarization of thermal emissions from dust grains.
Both methods are based on the fact that the dust grains are aligned along the magnetic field. In the former, background starlight is polarized parallel to the magnetic field.
Many previous observations have indicated that the global magnetic field is nearly perpendicular to the main massive filaments
\deleted{\citep{2011ApJ...734...63S,2011ApJ...741...21C,2016ApJ...830L..23K}}
\added{\citep{2011ApJ...734...63S,2011ApJ...741...21C,2013A&A...550A..38P,2016ApJ...830L..23K}}.

The far infrared- and millimeter-wave polarization observations assume that the thermal emission from magnetically aligned dust is polarized in the direction perpendicular to the interstellar magnetic field.
From the Planck all-sky survey, Planck Collaboration Int. XXXV
(\citeyear{2016A&A...586A.138P}) performed a statistical study for molecular clouds in the Gould belt to determine
 whether the interstellar magnetic field is parallel or perpendicular to the major axis of filamentary structures.
They found that the interstellar magnetic field is preferentially observed perpendicular to massive bright filaments, but filaments with low column density (striations) often extend in a direction parallel to the magnetic field. 
This trend is clearly seen in typical molecular clouds, such as Taurus, Lupus, and Chamaeleon-Musca.

Because the polarization is made by the magnetically aligned dust grains integrated along the line of sight, the apparent angle between the magnetic field and the filament is affected by their three-dimensional configuration (\citealp{2015ApJ...807...47T};
Planck Collaboration Int. XXXIII \citeyear{2016A&A...586A.136P}; \citealp{2020ApJ...899...28D}; \citealp{2020MNRAS.tmp.2934R}). 
For example, \citet{2020ApJ...899...28D} demonstrated that a major filament seems to extend parallel to the magnetic field observed in NGC 1333 by the James Clerk Maxwell Telescope, but the filament completely conforms to the commonly assumed configuration in which the magnetic field and filament are perpendicular to each other in three dimensions.

The stability of the interstellar filaments is often discussed using the mass per unit length, that is, the line-mass \citep{1963AcA....13...30S,1964ApJ...140.1056O}.  
In the case of an infinite cylindrical {\it isothermal} cloud with the central density $\rho_c$, the density profile $\rho(r)$ is given analytically as
\begin{equation}\label{eq:density_ost}
    \rho (r)=\rho_c\left(1+\frac{r^2}{8\ell^2}\right)^{-2},
\end{equation}
 where $\ell$ is the scale height that is expressed as $\ell=c_s/(4\pi G \rho_c)^{1/2}$ by using the isothermal sound speed $c_s$ and the gravitational constant $G$ \citep{1963AcA....13...30S,1964ApJ...140.1056O}.
Integration of Equation (\ref{eq:density_ost})  along the radius $r$ gives the line-mass of a filament with the surface radius $r_s$ as
\begin{equation}
    \lambda(r_s)=\int^{r_s}_{0}2\pi \rho r dr=\frac{2c^2_s}{G} \frac{r_s^2/8\ell^2}{1+r_s^2/8\ell^2}.
\end{equation}

The maximum line-mass that can be supported against self-gravity is given by $r_s/\ell\rightarrow \infty$ as $\lambda_{\rm iso, crit}=2c^2_s/G$, which is called the critical line-mass.
The critical line-mass is often used as a quantity
that controls star formation inside the molecular cloud \added{\citep{1987PThPh..77..635N,1992ApJ...388..392I,2014prpl.conf...27A}}.
When the line-mass $\lambda$ exceeds the critical line-mass $\lambda_{\rm iso,crit}$
in a filament, it contracts radially and begins star formation.
 For example, from the {\it Herschel} survey of the Aquila region, supercritical filaments with $\lambda>\lambda_{\rm iso,crit}$ contain most \deleted{($60\%$)} of the gravitationally bound prestellar cores \citep{2010A&A...518L.102A}.
 
Because star formation basically proceeds by gravitational contraction, knowing the conditions under which gravitational contraction begins leads to an understanding of the earliest phase of star formation. The equilibrium state of filaments has been studied from this standpoint.
\added{In addition to this, the magnetic field exists with the filamentary structures. The magnetic field is also playing a central role in star formation (for a review, see, for example, \citet{2019FrASS...6....5H}). Thus, to understand the effect of this magnetic field on the equilibrium state, we should study the magnetohydrostatic equilibrium state.}

\added{\citet{1963AcA....13...30S} studied  the equilibria of isothermal cylinders with the magnetic field parallel to their long axis.  When the plasma $\beta$ is constant, the critical line-mass increases owing to the increase of the scale-length $\ell$. 
In contrast, \citet{2000MNRAS.311...85F} found that the toroidal magnetic field has an opposite effect to compress the filaments and reduces the critical line-mass. 
However, these papers have discussed the case that the magnetic field is globally parallel to the filament.}

\citet{2014ApJ...785...24T} studied the magnetohydrostatic equilibrium state of a filamentary isothermal cloud threaded by a lateral magnetic field. The study assumed a magnetized infinitely long cylindrical isothermal cloud and studied the effect of the magnetic field for the maximum line-mass $\lambda_{\rm iso,max}$.
\citet{2014ApJ...785...24T} numerically derived an empirical formula of the maximum line-mass as
\begin{equation}
    \lambda_{\rm iso,max}\simeq 0.24\frac{\Phi_{\rm cl}}{G^{1/2}} +1.66\frac{c^2_s}{G},
\end{equation}
where $\Phi_{\rm cl}$ is one-half of the magnetic flux threading the filament per unit length. The study concluded that the maximum line-mass supported against self-gravity is represented by the function of the magnetic flux $\Phi_{\rm cl}$ and when considering a filamentary cloud, it is necessary to account for the magnetic field.

To characterize the density of an axisymmetric filament, Plummer-like profiles are often used as
    \begin{equation}\label{eq:plummer-like}
        \rho(r)=\frac{\rho_c}{[1+(r/R_f)^2]^{p/2}},
    \end{equation}
where $\rho_c$ is the central density, $R_f$ is the core radius, and $p$ is a density slope parameter \citep{2008MNRAS.384..755N,2011A&A...529L...6A}. 
The slope of the power-law distribution is determined by $p$, and the non-magnetized isothermal cylinder corresponds to $p=4$ [see Eq. (\ref{eq:density_ost})]. Density profiles observed by {\it Herschel} are well reproduced 
by a power-law distribution with an index around $p\simeq 2$.  For example, this index is $p\simeq 2.2 \pm 0.4$ for IC5146, $p\simeq 2.4 \pm 0.6$ for Aquila, and $p\simeq 2.3 \pm 0.1$ for Taurus \citep{2019A&A...621A..42A}.

\citet{2011ApJ...739L...2P}, \citet{2011A&A...533A..34H}, and \citet{2012ApJ...745..117B} reported that the observed filament profile was fitted well with the isothermal model  [Eq.(\ref{eq:density_ost})] rather than $p\simeq2$.
However, because these are based on observations with high-density tracers
 ($\rm NH_3,~ N_2H^+,~ and~ C^{18}O$), the obtained distribution may be affected by the abundance gradient. 
Even with the same {\it Herschel} data, \citet{2019MNRAS.489..962H} pointed out that the shallow radial density gradient $(p\simeq2)$ seems to be affected by the smoothing and averaging inherent in its derivation. In addition, they claimed that when analyzing each small local segment of the filament (of length $0.004$ pc), the data indicate rather $p=4$ than $p=2$.

Although more deliberation may be needed, so far, there is no strong evidence 
 to reject $p\simeq2$.
Thus, in this paper, we explore the physical reason why the density profile of the filament is fitted with $p\simeq 2$.
\deleted{To explain a shallow density gradient like $p \simeq 2$, \citet{2015MNRAS.446.2110T} proposed the effect of the non-magnetized gas obeying a non-isothermal equation of state. }
\added{To explain a shallow density gradient like $p \simeq 2$, \citet{2015MNRAS.446.2110T} assumed a filament supported by thermal and non-thermal motions then proposed the effect of the non-magnetized gas obeying a non-isothermal equation of state.}

Additionally, \citet{2019A&A...621A..42A} and \citet{2019MNRAS.489..962H} reported that the temperature at the filament center is lower than that at the surface; that is, the filament has a negative temperature gradient. \added{This seems to be explained as the central part is shielded from the incoming interstellar radiation by the outer layer.}

Based on this, \citet{2015MNRAS.446.2110T} studied the non-magnetized infinite cylinder obeying the non-isothermal polytropic equation of state as follows:
\begin{equation}\label{eq:polytropic}
    p_g=K\rho^\gamma,
\end{equation}
where $\gamma$, $p_g$, $K$, and $\rho$ are the polytropic exponent, gas pressure, proportional constant, and gas density, respectively.
The polytropic exponent is often used as $\gamma=1+1/N$, where $N$ is the polytropic index.
The polytropic index $N$ represents how steep the temperature gradient is, and is negative ($N\le-1$) when the filament has a negative temperature gradient.
\citet{2015MNRAS.446.2110T} concluded that a negative polytropic index ($-\infty < N < -1$) makes the density profile shallower than that of an isothermal model, $N=-\infty$ (see Fig.~1 of their paper).  
\added{\citet{2015MNRAS.446.2118T} also studied the effect of a helical magnetic field on the polytropic filament.
They reported the pitch angle, which is the angle between poloidal and toroidal magnetic fields, determines whether the magnetic field compresses or supports the polytropic filaments similar to the isothermal ones \citep{2000MNRAS.311...85F}.
}

In this paper, we present the numerical calculation for the equilibrium state of a magnetized filament that has a negative temperature gradient. 
The structure of this paper is as follows.
In section \ref{sec:method}, we introduce the model and formulation for this calculation.
We show the numerical result for this filament in section \ref{sec:result}. 
In section \ref{sec:discussion}, we discuss the effects of the magnetic field and the negative temperature gradient on the line-mass and the filament structure. 
We summarize  the results of this paper and provide conclusions in section \ref{sec:summary}.

\section{Method}\label{sec:method}
The method to obtain the magnetohydrostatic structure for a polytropic gas is formulated based on the method for an isothermal gas \citep{2014ApJ...785...24T}.
\subsection{Basic Equations}
To derive the magnetohydrostatic configuration, we start from the following four equations. 
First, the polytropic equation is 
\begin{equation}\label{eq:polytropic_equation}
    p_g=K\rho^{1+1/N},
\end{equation}
in which the meaning of the variables is the same as in Equation (\ref{eq:polytropic}). 
The equation is based on the assumption that the pressure and the density are connected with the polytropic index. 
The second equation is a force balance equation between the Lorentz force, gravitational force, and pressure gradient, which is written as
    \begin{equation}\label{eq:balance}
        \frac{1}{c}{\bf j}\times {\bf B}- \rho \nabla \psi -\nabla p_g=0,
    \end{equation}
where $c,~{\bf j},~\rm \bf B$, and $\psi$ represent the light speed, electric current density, magnetic flux density, and gravitational potential, respectively. 
The third equation is Poisson's equation for self-gravity, which is expressed as
     \begin{equation}\label{eq:poasson_psi}
         \nabla^2 \psi=4\pi G \rho,
     \end{equation}
     where $G$ is the gravitational constant.
     The fourth equation is Amp\`{e}re's law, which is written as \begin{equation}
     \label{eq:ampere}
         {\bf j}=\frac{c}{4\pi}\nabla \times {\bf B}.
     \end{equation}
     We search for a solution for a filament extending infinitely in the $z$-direction and assume all the physical quantities depend on only $(x,y)$. 
     We introduce a magnetic flux function $\Phi$, from which the magnetic flux density ${\bf B}=(B_x,B_y)$ is given as
     \begin{subequations}
     \begin{align}
      &{ B}_x=-\frac{\partial {\Phi}}{\partial y},\\
      &{ B}_y=\frac{\partial {\Phi}}{\partial x}.
      \end{align}
     \end{subequations}
     It is noted that in two dimensions, the magnetic field line is given by a contour line of $\Phi(x,y)=\rm const$.
     From Equation (\ref{eq:ampere}), the electric current $\bf j$ is rewritten as
     \begin{subequations}
     \begin{align}
             &{ j}_x=\frac{c}{4\pi}\frac{\partial { B}_z}{\partial y},\\
             &{ j}_y=-\frac{c}{4\pi}\frac{\partial { B}_z}{\partial x},\\
             &{ j}_z=\frac{c}{4\pi}\left(\frac{\partial B_y}{\partial x}-\frac{\partial B_x}{\partial y}\right)=\frac{c}{4\pi}\Delta_2 \Phi,
     \end{align}
     \end{subequations}
     where $\Delta_2$ is defined as
     \begin{equation}
         \Delta_2 \equiv \nabla^2_2=\partial^2 /\partial x^2+ \partial^2/\partial y^2.
     \end{equation}
     Hereafter, $\nabla_2 \equiv (\partial/\partial x, \partial/\partial y)$ represents the two-dimensional differentiation operator. 
     For the polytropic gas, the pressure term of Equation (\ref{eq:balance}) is
     \begin{equation}
         \frac{1}{\rho}\nabla_2 p_g=\frac{K}{\rho}\nabla_2 \rho^{1+\frac{1}{N}}=K(N+1)\nabla_2 \rho^{1/N}.
     \end{equation}
     Using this equation, the force balance Equation (\ref{eq:balance}) becomes
     \begin{subequations}
     \begin{align}
        & x\mbox{-component}:-\frac{1}{4\pi}\Delta_2\Phi\frac{\partial \Phi}{\partial x}-\rho \frac{\partial \psi}{\partial x}- K(N+1)\rho\frac{\partial \rho^{1/N}}{\partial x}-\frac{1}{8 \pi}\frac{\partial B_z^2}{\partial x}=0,\\
        & y\mbox{-component}: -\frac{1}{4\pi}\Delta_2\Phi\frac{\partial \Phi}{\partial y}-\rho \frac{\partial \psi}{\partial y}-K(N+1)\rho\frac{\partial \rho^{1/N}}{\partial y}-\frac{1}{8 \pi}\frac{\partial B_z^2}{\partial y}=0,
    \end{align}
     \end{subequations}
     and when $B_z=0$, these two equations reduce to
    \begin{subequations}\label{eq:poa}
     \begin{align}
        &x\mbox{-component}: -\frac{1}{4\pi}\Delta_2\Phi\frac{\partial \Phi}{\partial x}-\rho \frac{\partial }{\partial x}\left[ \psi +K(N+1)\rho^{{1}/{N}}\right]=0,\\
        & y\mbox{-component}:-\frac{1}{4\pi}\Delta_2\Phi\frac{\partial \Phi}{\partial y}-\rho \frac{\partial }{\partial y}\left[ \psi +K(N+1)\rho^{{1}/{N}}\right]=0.
     \end{align} 
     \end{subequations}
     By taking the inner product of Equation (\ref{eq:poa}) and ${\rm{\bf B}}=(B_x,B_y)$,
     \begin{equation}
         \rho ~({\bf B \cdot \nabla_2})\left[ \psi +K(N+1)\rho^{{1}/{N}}\right]=0
     \end{equation}
     is required in the direction parallel to the magnetic field lines. This means the quantity $\psi+K(N+1)\rho^{1/N}$ is constant along a magnetic flux tube as
     \begin{equation}\label{eq:Ber}
          \psi +K(N+1)\rho^{{1}/{N}}=H(\Phi),
     \end{equation}
     where $H(\Phi)$ is the Bernoulli constant, which is a function dependent only on $\Phi$. Along a magnetic tube given by a constant $\Phi$, the density is calculated from the gravitational potential
     \begin{equation}\label{eq:density}
         \rho=\left[\frac{H(\Phi)-\psi}{K(N+1)}\right]^N,
     \end{equation}
     and then Equation (\ref{eq:poa}) is rewritten as
     \begin{subequations} \label{eq:poasson_phi}
     \begin{align}
     & -\frac{1}{4\pi}\Delta_2\Phi\frac{\partial \Phi}{\partial x}=\rho\frac{\partial H}{\partial x}=\rho \frac{d H}{d\Phi}\frac{\partial \Phi}{\partial x},\\
     & -\frac{1}{4\pi}\Delta_2\Phi\frac{\partial \Phi}{\partial y}=\rho\frac{\partial H}{\partial y}=\rho \frac{d H}{d\Phi}\frac{\partial \Phi}{\partial y}.
     \end{align}
     \end{subequations} 
Similar to the method in \citet{2014ApJ...785...24T}, we assume that forces are balanced inside the filament [Eq.\,(\ref{eq:poasson_phi})].\added{Outside the filament, we assume that a force-free magnetic field (the electric current ${\bf j}=0$) and hot tenuous gas ($\rho=0$) exist. The extended gas confines the filament with its pressure (external pressure $p_{\rm ext}$).} \deleted{Outside the filament, however, a tenuous and hot medium ($\rho=0$) is extended and confines the filament with its pressure.}
Thus, the right-hand side of Equation (\ref{eq:poasson_phi}) vanishes outside the filament.
Finally, we can derive the two basic equations. Equation (\ref{eq:poasson_phi}) leads to
     \begin{equation}
     \quad -\Delta_2\Phi =
        \left\{
         \begin{array}{cl}
              4\pi \rho\frac{dH}{d\Phi}&({\rm inside~the~ filament}),\\ 0&({\rm outside~the~ filament}), 
         \end{array}
         \right.
     \end{equation} 
 and with use of Equation (\ref{eq:density}), Poisson's equation (\ref{eq:poasson_psi}) is rewritten as 
     \begin{equation}
      \quad \Delta_2\psi =
      \left\{
         \begin{array}{cl}
         4\pi G\left[\frac{H(\Phi)-\psi}{K(N+1)}\right]^N&({\rm inside~the ~filament}),\\
         0&({\rm outside~ the ~filament}).
         \end{array}
         \right.
     \end{equation}

We find the equilibrium state by solving these two second-order differential equations simultaneously by the self-consistent field method, but we need to know the value of $H(\Phi)$ at each magnetic field line. 
\subsection{Mass Loading}
Here, we introduce a mechanism to derive $H(\Phi)$. In this paper, we assume that a large-scale magnetic field runs along the $y$-direction. 
We assume vertical symmetry at $y=0$. 
Then a line-mass $\Delta \lambda$ that is contained between two magnetic field lines, $\Phi$ and $\Phi+\Delta \Phi$, is expressed as
    \begin{subequations}
    \begin{align}
            \Delta \lambda(\Phi)&=2\int^{y_s(\Phi)}_{0}\int^{x(y,\Phi+\Delta\Phi)}_{x(y,\Phi)}\rho(x,y)dxdy\\
            &=2\int^{y_s(\Phi)}_{0}\frac{\rho}{(\partial \Phi/\partial x)}\Delta\Phi dy\\
            &=2\int^{y_s(\Phi)}_{0}\left[\frac{H(\Phi)-\psi}{K(N+1)} \right]^N (\partial \Phi/\partial x)^{-1}\Delta \Phi dy,
    \end{align}        
    \end{subequations}
    where $y_s$ represents the $y$-coordinate of the filament surface and $H(\Phi)$ stays constant in the integration. 
    It leads to a problem of finding an appropriate set of $H(\Phi)$ and $y_s(\Phi)$ at the same time satisfying the  following equation
    \begin{equation} \label{eq:mass_loading}
        \frac{d\lambda}{d\Phi}=2\int^{y_s(\Phi)}_{0}\left[\frac{H(\Phi)-\psi}{K(N+1)} \right]^N (\partial \Phi/\partial x)^{-1}dy. 
    \end{equation}
    The left side of this equation is given as a model of the mass-to-flux ratio distribution, which is called mass-loading. 
    
\begin{figure}
    \centering
    \includegraphics[keepaspectratio,scale=0.5]{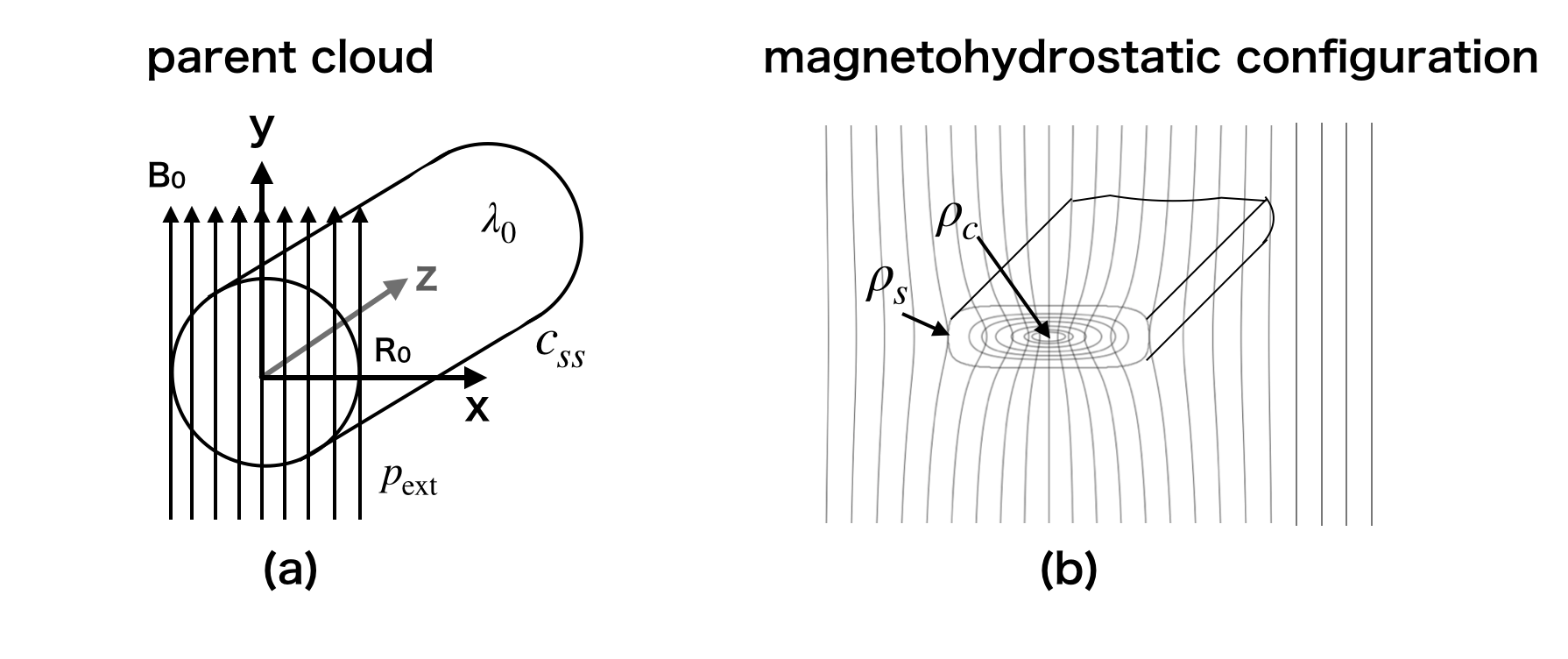}
      \caption{\added{Model. The parent cloud (a) has uniform density with radius $R_0$, which is threaded with the uniform magnetic field $B_0$. 
      The parent cloud is immersed in the external pressure $p_{\rm ext}$ and given the line-mass of $\lambda_0$.  
      Starting from this state we search the equilibrium state (b) with assuming the flux freezing that keeps the initial mass distribution against the magnetic flux. 
      Gas obeys the polytropic equation $p_g=K\rho^{1+1/N}$.
      The polytropic index $N$, the radius of the parent cloud $R'_0\equiv R_0/[c_{\rm ss}/(4\pi G\rho_s)^{1/2}]$, the plasma beta $\beta_0\equiv p_{\rm ext}/(B^2_0/8\pi)$, and the line-mass $\lambda'\equiv \lambda_0/(c^2_{\rm ss}/4\pi G)$ determine the equilibrium state.
      It is noted that we use the center-to-surface density ratio $\rho'_c\equiv \rho_c/\rho_s$ as the fourth parameter instead of the line-mass $\lambda'$ because it is easier to find the equilibrium state.
      }}
\label{fig:model}
\end{figure}

    We assume the central density of the filament as $\rho_c$ and the potential as $\psi_c$. Then, Equation (\ref{eq:Ber}) gives the value of the Bernoulli constant for the central magnetic field line coinciding with the $y$-axis, which is specified by $\Phi=0$, as 
    \begin{equation}\label{eq:H_0}
        H(\Phi=0)=K(N+1)\rho_c^{1/N}+\psi_c.
    \end{equation}
    Equation (\ref{eq:mass_loading}) uses Equation (\ref{eq:H_0}) to obtain the mass-loading on the central magnetic field line $\Phi=0$ as follows:
    \begin{equation}\label{eq:phi_psi}
        \left. \frac{d\lambda}{d\Phi}\right|_{\Phi=0}=2\int^{y_s}_{0}\left[\rho_c^{1/N}+\frac{\psi_c-\psi}{K(N+1)}\right]^N(\partial \Phi/\partial x)^{-1}dy,
    \end{equation}
    where the upper boundary $y_s$ is given as a point where $\psi=\psi_s$.
    The surface potential is written as
    \begin{equation}
        \psi_s=\psi_c+K(N+1)(\rho_c^{1/N}-\rho_s^{1/N}),
    \end{equation}
    \added{where the $\rho_s$ is the density at the filament surface in the equilibrium state (see Figure \ref{fig:model} (b)).}
    Equation (\ref{eq:phi_psi}) indicates that, from a set of potentials $\Phi$ and $\psi$, the mass-loading for the central magnetic field line $\left. d\lambda/d\Phi\right|_{\Phi=0}$ is obtained as a function of $\rho_c$.

\deleted{In this paper, we assume the following mass-loading distribution:}
\added{In this paper, for comparing with the isothermal model, we assume the following mass-loading distribution: }
    \begin{equation}
        \frac{d\lambda}{d\Phi}=\left. \frac{d\lambda}{d\Phi}\right|_{\Phi=0}\left[1-\left(\frac{\Phi}{\Phi_{\rm cl}}\right)^2\right]^{1/2},
    \end{equation}
    \added{which is the same assumed in \citet{2014ApJ...785...24T}.}
In this equation, $\Phi_{\rm cl}$ represents the magnetic flux threading the unit length of the filament and $\Phi_{\rm cl}= R_0\cdot B_0$, where $R_0$ and $B_0$ represent the initial radius of the filament and magnetic field strength, respectively, of the initial uniform magnetic field. 
This is realized when a uniform-density cylindrical filament is threaded with a uniform magnetic field, where $-\Phi_{\rm cl} \le \Phi \le \Phi_{\rm cl}$ represents the magnetic field line threading the filament, while $\Phi<-\Phi_{\rm cl}$ and $\Phi>\Phi_{\rm cl}$ represent the magnetic field lines not threading the filament \added{(see Figure \ref{fig:model})}.
In this paper, we call this filament with uniform density and uniform magnetic field as the ``parent'' cloud, which gives the mass-loading in the filament in equilibrium.     
That is, we assume that the mass loading is determined by the parent cloud and is conserved by flux freezing.
    
    For the magnetic field lines $\Phi \neq 0$, the equation becomes
    \begin{eqnarray}
        \left. \frac{d \lambda}{d \Phi}\right|_{\Phi=0}\left[1-\left(\frac{\Phi}{\Phi_{\rm cl}}\right)^2\right]^{1/2}
        &=&2\int^{y_s(\Phi)}_{0}\left[\rho_s^{1/N}+\frac{\psi_s(x(\Phi,y_s),y_s)-\psi(x(\Phi,y),y)}{K(N+1)}\right]^N (\partial \Phi/\partial x)^{-1}dy.
    \end{eqnarray}
    For a given $\Phi$, $y_s(\Phi)$ is chosen to satisfy the above equation, which is achieved with use of the bisection method of non-linear equations. 
    
  \subsection{Normalization Units}\label{sec:2.3}
     \begin{table*}
    \caption{Units used for normalization.  \deleted{It should be noted that physical variables are normalized with their quantities at the filament surface.}\added{Physical variables are normalized with their quantities at the filament surface.} Because the free-fall time scale ($t_{\rm ff}$) and scale length ($L$) are defined on the filament surface, the unit of speed is taken as the isothermal sound speed ($c_{\rm ss}$) at the surface of the filament.}          
    \label{tab:normalization}      
    \centering          
    \begin{tabular}{ll}
    \hline                    
      Unit of pressure\dotfill  & External pressure, $p_{\rm ext}$ \\
        Unit of density\dotfill &  Density at the surface, $\rho_s$ \\
        Unit of time\dotfill &  Free-fall time, $t_{\rm ff}=(4\pi G\rho_s)^{-1/2}$ \\ 
        Unit of speed\dotfill & $c_{\rm ss}=(p_{\rm ext}/\rho_s)^{1/2}=(K\rho_s^{1/N})^{1/2}$  \\
        Unit of magnetic field strength\dotfill & $B_u=(8\pi p_{\rm ext})^{1/2}$ \\
        Unit of length\dotfill & $L=c_{\rm ss}t_{\rm ff}=c_{\rm ss}/(4\pi G\rho_s)^{1/2}=(K\rho_s^{-1+1/N}/4\pi G)^{1/2}$\\
        Unit of temperature\dotfill  & Temperature at the surface, $T_s$\\
    \hline                  
    \end{tabular}
    \end{table*}

    \added{In this paper, physical variables are normalized with their quantities at the filament surface. We regard the following three quantities fundamental: the external pressure $p_{\rm ext}$, the surface density $\rho_s$, and the isothermal sound speed $c_{\rm ss}$ at the filament surface.
    Then, for example, the scale length ($L$) is defined by the free-fall time ($t_{\rm ff}$) at the filament surface density ($\rho_s$) and the isothermal sound speed ($c_{\rm ss}$) at the surface of the filament, as $L=c_{\rm ss}t_{\rm ff}$.}
    From the polytropic equation, the external pressure $p_{\rm ext}$ and the surface density $\rho_s$ are related as 
    \begin{equation}
        p_{\rm ext}=K\rho_s^{1+1/N}.
    \end{equation}
    \deleted{Then, we use the scales characterizing the system shown in Table \ref{tab:normalization} to define normalized variables as}
    \added{The physical scales characterizing the system are given as in Table \ref{tab:normalization}. We define the normalized variables as}
    \begin{subequations}
        \begin{align}
            &p'\equiv  p_g/p_{\rm ext},\\
            &\rho'\equiv \rho/\rho_s,\\
            &\psi'\equiv\psi/c_{\rm ss}^2=\psi/(K\rho_s^{1/N}),\\
            &\Phi'\equiv\Phi/(B_uL),\\
            &x'\equiv x/L,\\
            &y'\equiv y/L,\\
            &\Delta'_2\equiv \Delta_2\cdot L^2,\\
            &\lambda'\equiv \lambda/(\rho_s L^2).
        \end{align}   
    \end{subequations}
    \deleted{where the $c_{\rm ss}$ is the isothermal sound speed at the surface of the filament.}
    \added{where the prime represents the normalized variables.}
    The density is normalized as
    \begin{equation}
        \rho'=\left[\frac{H'-\psi'}{N+1}\right]^N.
    \end{equation}
   Poisson's equation is rewritten as 
   \begin{equation}
      \Delta'_2\psi'=\rho'=\left[\frac{H'(\Phi')-\psi'}{N+1}\right]^N,
   \end{equation}
   while Poisson's equation for magnetic flux function is given as 
   \begin{equation}
       -\Delta'_2\Phi'=\frac{1}{2}\rho'\frac{dH'}{d\Phi'}.
   \end{equation}
   The mass-loading on the central magnetic field is given as 
   \begin{equation}
      \left. \frac{d\lambda'}{d\Phi'}\right|_{\Phi=0} =2\int^{y'_s}_{0} \left[{\rho'_c}^{1/N} +\frac{\psi'_c-\psi'}{N+1}\right]^N {\left(\frac{\partial \Phi'}{\partial x'}\right)^{-1}}dy', 
   \end{equation}
   and Equation (\ref{eq:mass_loading}) reduces to
   \begin{equation}
       \frac{d\lambda'}{d\Phi'}=2\int^{y_s'}_{0}\left[1+\frac{\psi'_s-\psi'}{N+1}\right]^N\left(\frac{\partial \Phi'}{\partial x'}\right)^{-1}dy'.
   \end{equation}
   
   The two Poisson equations require boundary conditions. We impose the Dirichlet boundary condition on the outer numerical boundary given below.
   Far from the origin, we assume that the gravitational potential $\psi$ converges to that realized for a line-mass $\lambda$ placed at the origin as 
   \begin{equation}\label{eq:boundary_psi}
       \psi=2G\lambda {\rm log}r.
   \end{equation}
     The outer boundary condition for the magnetic potential is expressed as
   \begin{equation}
    \Phi=B_0 x,  
   \end{equation}
   in which we assume that the magnetic field is connected to the uniform magnetic field with strength $B_0$ far from the center. Then we normalize the two potentials. The normalized value $\lambda'=\lambda/ \rho_s L^2$ is used to reduce Equation (\ref{eq:boundary_psi}) to
   \begin{equation}\label{eq:grav_pot_non_dim}
       \psi'=\frac{\lambda'}{2\pi}{\rm log}r'.
   \end{equation}
   The line-mass $\lambda$ is given as 
   \begin{subequations}\label{eq:lambda_0}
   \begin{align}
     \lambda&=2\int^{\Phi_{\rm cl}}_{0}\frac{d\lambda}{d\Phi}d\Phi\\
     &=2\left. \frac{d\lambda}{d\Phi}\right|_{\Phi=0}\int^{\Phi_{\rm cl}}_{0}\left[1-\left(\frac{\Phi}{\Phi_{\rm cl}}\right)^2\right]^{1/2}d\Phi\\
     &=\frac{\pi}{2}\Phi_{\rm cl}\left. \frac{d\lambda}{d\Phi}\right|_{\Phi=0}.
    \end{align} 
   \end{subequations}
   If we know the mass-to-magnetic flux ratio at the center, $\left. {d\lambda}/{d\Phi}\right|_{\Phi=0}$, we obtain the boundary value of $\psi'$ after calculating $\lambda$ using Equations (\ref{eq:grav_pot_non_dim}) and (\ref{eq:lambda_0}). 
 The magnetic field potential is normalized as
   \begin{equation}
       \Phi'=\beta_0^{-1/2}x',
   \end{equation}
   where $\beta_0$ is a ratio of the external pressure $p_{\rm ext}$ to the magnetic pressure $B_0^2/8\pi$ and defined as
   \begin{equation}
       \beta_0 \equiv \frac{p_{\rm ext}}{B^2_0/8\pi}.
   \end{equation}
   
   \subsection{Parameters}
   After the normalization, a solution is specified by four non-dimensional parameters, $\Phi'_{\rm cl}$ , $\beta_0$ , \deleted{$\rho'_c$}\added{$\rho'_c\equiv\rho_c/\rho_s$} , and $N$\added{(see Figure \ref{fig:model})}.  
   The non-dimensional magnetic flux $\Phi'_{\rm cl}$ is given as
   \begin{equation}
       \Phi'_{\rm cl} =R'_{0}\cdot B'_0=R'_{0}\cdot \beta^{-1/2}_0,
   \end{equation}
   where $R'_0$, which is defined as the initial radius of uniform filament $R_0$ normalized by the scale length $L$. Hereafter, we omit the prime, which indicates normalized quantities, unless the meaning is unclear. 
   
    \begin{table}
  \begin{center}
  \caption{Model parameters and maximum supported line-mass.
  The column $\rho_{c\ \rm Max}$ indicates that the solutions are obtained between $\rho_c=2$ and $\rho_{c\ \rm Max}$.
  Numbers marked with the symbol * represent lower limits of $\lambda_{\rm max}$.
  \deleted{The symbols $\dagger$ and $\diamond$ represent a grid spacing of 0.1/16 and 0.1/8, respectively, and in both models the number of grid points is $1281\times1281$.}
  \added{The symbols $\dagger$, $\diamond$, and $\bullet$ represent a grid spacing of 0.1/16, 0.1/8, and 0.1/4, respectively, and in the models with $\dagger$ and $\diamond$ the number of grid points is $1281\times1281$ and that of the model with $\bullet$ is $641\times641$.}
  The rest of the models are calculated with grid points of $641\times641$ and a grid spacing of $0.1/8$.
  }
  \label{tab:tab1}
  \begin{tabular}{lccccccccccc} \hline\hline
  {Model} & $R_0$ & $\beta_0$ & \multicolumn{4}{c}{$\rho_{c\ \rm Max}$} & $\Phi_{\rm cl}$ &
  \multicolumn{4}{c}{$\lambda_{\rm max}$}\\ \cline{4-7}\cline{9-12}
  & & & $N=-3$ & $-5$ & $-10$ & $-100$ & & $N=-3$ & $-5$ & $-10$ & $-100$\\ \hline
    R1$\beta$1  & 1& 1  & 500 & $10^3$ & $10^3$ & 200    & 1   & 9.957& 13.42& 17.54& 23.90\\
    R1$\beta$0.5& 1& 0.5& 500 & $10^3$ & $10^3$ & 500    & 1.41& 11.06& 14.46& 18.84& 25.60\\
    R1$\beta$0.1& 1& 0.1& 500 & $10^3$ & $10^3$ & $10^3$ & 3.16& 16.30& 20.51& 25.31& 32.52\\
    R1$\beta$0.05& 1& 0.05 & 500 & $10^3$ & $10^3$ & $10^3$ & 4.47&19.83*& 25.27& 30.54& 38.03\\
    R2$\beta$1& 2& 1 &     $10^3$ & $10^3$ & $10^3$ & 100 & 2&14.36& 17.77& 21.57& 26.84\\
    R2$\beta$0.5& 2& 0.5 & $10^{3~\dagger}$ & $10^3$ & $10^3$ & 500 & 2.83& 17.22& 20.94& 24.96& 31.08\\
    R2$\beta$0.1& 2& 0.1 & $10^{3~\dagger}$ & $10^3$ & $10^3$ & $10^3$ & 6.32& 29.91& 34.63& 39.81& 47.02\\
    R2$\beta$0.05& 2& 0.05 & $10^{3~\dagger}$ & $10^3$ & $10^3$ & $10^3$ & 8.94& 38.78*& 44.75& 50.86& 58.74\\
    R5$\beta$1 & 5 & 1  & $10^{3~\diamond}$ & $10^{3~\bullet}$ & $10^{3~\bullet}$ & $10^{3~\bullet}$ & 5 & 27.72 & 31.48 & 35.32 & 40.98\\
    R5$\beta$0.5& 5& 0.5 &  $500^{~\diamond}$ & $10^{3~\bullet}$ & $10^{3~\bullet}$ & $10^{3~\bullet}$ & 7.07 & 35.90& 40.22& 44.64& 50.74\\
    R5$\beta$0.1& 5& 0.1 &  $500^{~\diamond}$ & $10^{3~\bullet}$ & $10^{3~\bullet}$ & $10^{3~\bullet}$ & 15.8 & 68.61& 76.14& 82.59& 91.37\\
    R5$\beta$0.05& 5& 0.05 & $500^{~\diamond}$ & $10^{3~\bullet}$ & $10^{3~\bullet}$ & $10^{3~\bullet}$ & 22.4 & 89.18*& 99.81*& 110.6& 120.5\\
    \hline
  \end{tabular}
  \end{center}
 \end{table}
\subsection{Numerical Method}
We solved Poisson's equation with the conjugate gradient
 method preconditioned with incomplete Cholesky  
 factorization (ICCG). 
The number of grid points was chosen as $641\times 641$ or $1281\times1281$ and the grid spacing was chosen $\Delta x=\Delta y=0.1/16$, $0.1/8$, or $0.1/4$.
\added{The outer numerical boundaries are placed at $x=y=\pm4$ in the models with $R_0=1$ and $2$ and at $x=y=\pm8$ in those with $R_0=5$.}
We summarize the model parameters in Table \ref{tab:tab1}.

We verified our calculation by solving an approximate isothermal equilibrium state with the polytropic method and assuming $N=-100$ and $-1000$.
From Equation (\ref{eq:polytropic}), the polytropic indices $N=-100$ and $N=-1000$ correspond to the polytropic exponents $\gamma~(N=-100)=0.99$ and $\gamma~(N=-1000)=0.999$, both of which are close to the isothermal case of $\gamma=1$.
We compare the equilibrium state of the isothermal  \citep{2014ApJ...785...24T} and polytropic filaments $(N=-100~{\rm and} -1000)$ while paying attention to the line-mass.
In this comparison, the other parameters, that is, the radius of the parent cloud $R_0=2$ and the plasma beta $\beta_0=0.1$, are constant.  
When the central density is $\rho_c=10^3$, the line-mass of the polytropic filament is $\lambda_{N=-100}=47.02$ and $\lambda_{N=-1000}=47.96$, while the isothermal one is $\lambda_{\rm iso}=48.07$.
When the central density is $\rho_c=10^2$, the corresponding line-masses are $\lambda_{N=-100}=43.31$, $\lambda_{N=-1000}=44.00$, and $\lambda_{\rm iso}=44.08$. 

The line-mass of polytropic filaments is slightly lower than the isothermal one. However, it is clearly shown that the line-mass converges to the isothermal value when $N$ moves to $-\infty$.
Thus, our calculation reproduces a line-mass close to the isothermal one when the polytropic index $N\rightarrow -\infty$. Hereafter, we assume the results with $N=-100$ as the isothermal model.

\section{Results}\label{sec:result}

\subsection{Comparison of Polytropic
{\rm({\it N} = --3)} and Isothermal {\rm({\it N} = --100)} Filaments }\label{sec:3.1}
In this section, we compare the density profile and the line-mass of the polytropic ($N=-3$) and isothermal ($N=-100$) filaments. In the comparison, other parameters of these filaments, the radius of the parent cloud $R_0=1$ and the plasma beta $\beta_0=0.1$, are constant.

\begin{figure}
     \begin{tabular}{ccc}
         \includegraphics[keepaspectratio,scale=0.4]{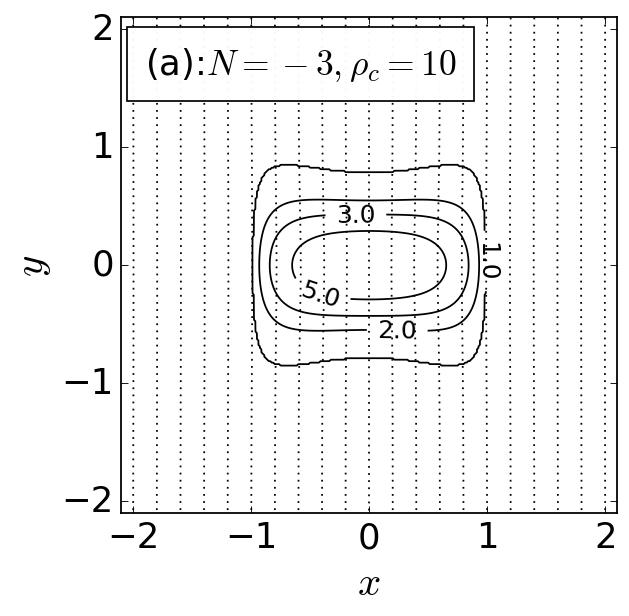} & \includegraphics[keepaspectratio,scale=0.4]{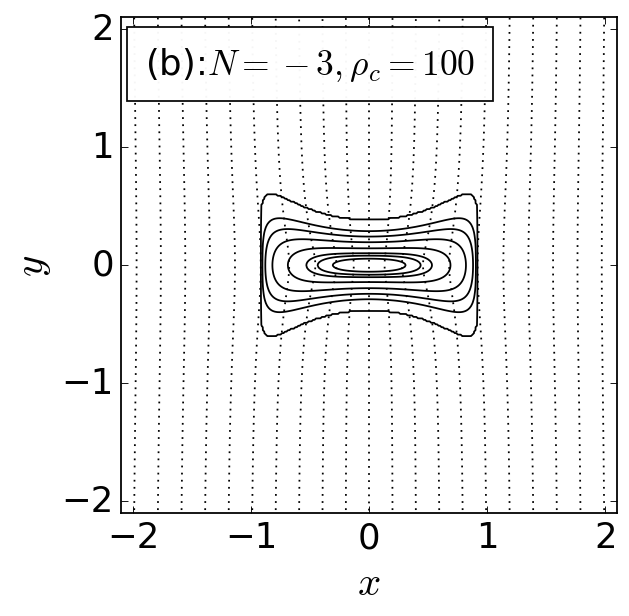}& \includegraphics[keepaspectratio,scale=0.4]{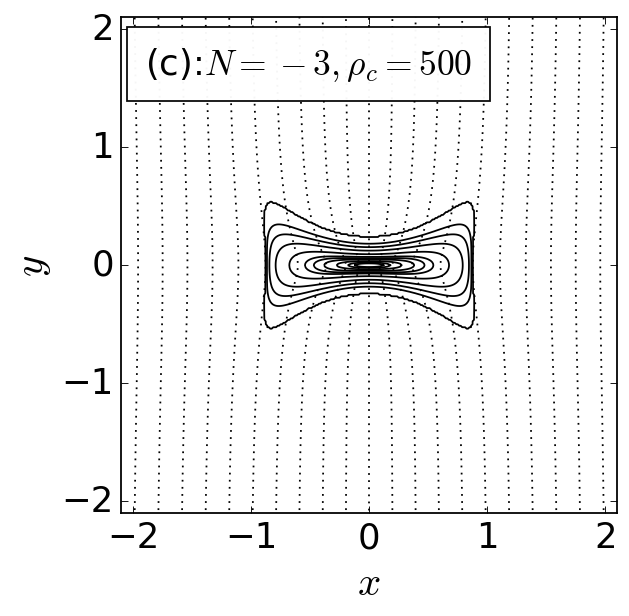}\\ \includegraphics[keepaspectratio,scale=0.4]{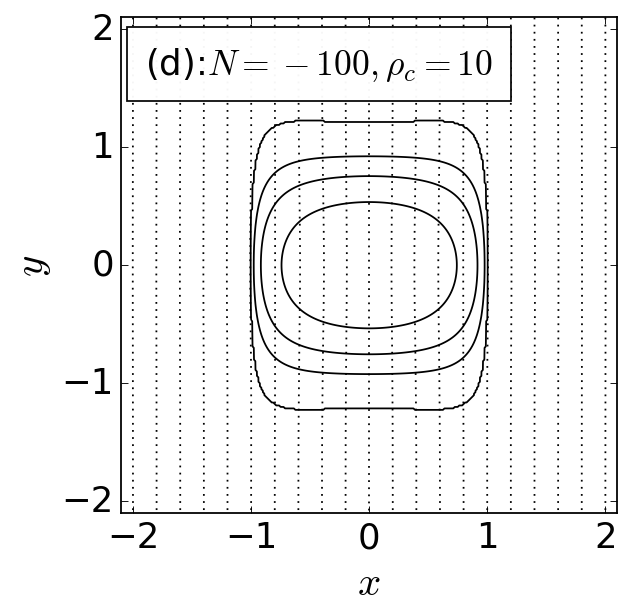}& \includegraphics[keepaspectratio,scale=0.4]{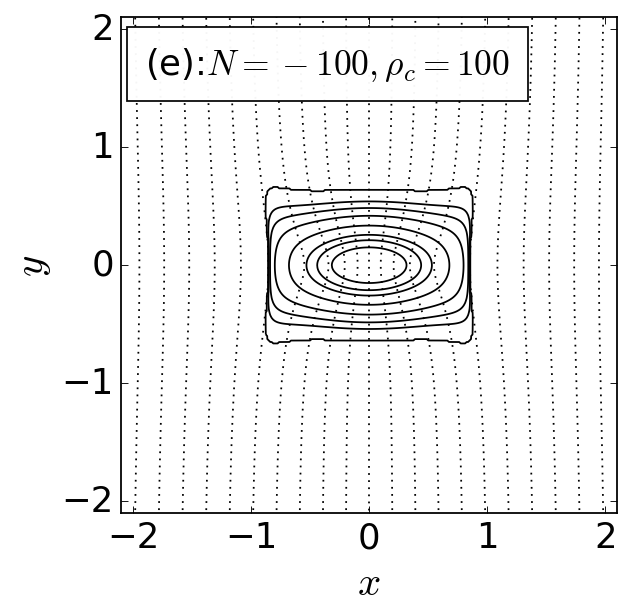}&\includegraphics[keepaspectratio,scale=0.4]{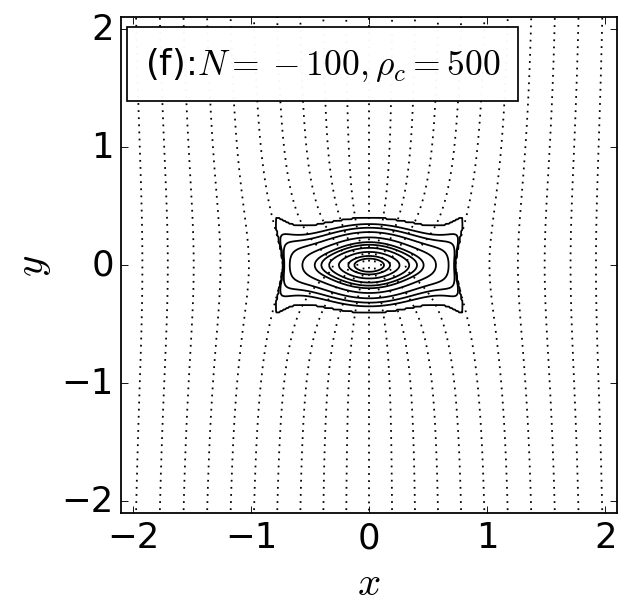} \\
     \end{tabular}
      \caption{Comparison of polytropic (a-c: $N=-3$) and isothermal (d-f: $N=-100$) filaments. This figure shows the cross section of the equilibrium state on the $x-y$ plane. Other parameters are constant: radius of the parent cloud $R_0=1$ and plasma beta $\beta_0=0.1$. We show three models with different central densities $\rho_c$ as $\rho_c=10$ [(a) and (d)], $\rho_c=100$ [(b) and (e)], and $\rho_c=500$ [(c) and (f)]. Solid lines show the isodensity contours, and each contour level is chosen as $\rho=2$, 3, 5, 10, 20, 30, 50, 100, 200, 300, 500, and 1000, respectively, from outside to inside.  Dotted vertical lines are magnetic field lines. The line-masses of these filaments are $\lambda_{N=-3}=10.36~{\rm (a)}, ~15.10~{\rm (b)}, ~16.30~{\rm (c)}$ and $\lambda_{N=-100}=17.90~{\rm (d)}, ~29.47~{\rm (e)}, ~32.18~{\rm (f)}$. }
\label{fig:cross_section_R1}
\end{figure}

First, we begin with the density distribution of the equilibrium state. 
We show the cross sections of the polytropic filaments in Figure \ref{fig:cross_section_R1} (a)-(c) and the isothermal filaments in (d)-(f).
In this paper, we call the $x$- and $y$-coordinates of the cloud surface crossing the $x$- and $y$-axes as the ``width'' $x_s$ and ``height'' $y_s$ of the filament, respectively.
It is shown that both of these filaments become flatter as the central density increases: the height of the filament shrinks, while the width remains almost unchanged. 
This is due to the character of the Lorentz force, which works in the perpendicular direction to the magnetic field line but does not work in the parallel direction. 
Because extra force to support the filament is working in the direction perpendicular to the magnetic field, the isodensity contours of the cross section shrink mainly in the $y$-direction and appear flat.

Figure \ref{fig:cross_section_R1} shows that the cross section of the polytropic filament is flatter than the isothermal one when two with the same central density are compared. 
However, the outer part's gas scale height in the $y$-direction of the polytropic filament is nearly equal to that of the isothermal one. 
For example, panel (b) shows that the polytropic filament has a height of $y_s\simeq 0.39$ on the symmetric $y$-axis. 
However, the surface inflates outwardly, and the height of the surface reaches $\simeq 0.60$ near $x\simeq 0.80$.
Thus, this polytropic filament has a maximum height of $\sim0.60$.
In contrast, the corresponding isothermal model [panel (e)] does not show such inflation  ($y_s\simeq 0.64$ and maximum height $\simeq 0.66$).
As is shown, the maximum height of the polytropic filament is nearly the same as that of the  isothermal filament.
This is understood by the temperature near the surface of the polytropic filament, which is not very different from the temperature of the isothermal filament, although the polytropic filament has a lower central temperature compared with the isothermal filament. 

\begin{figure}
     \begin{tabular}{ll}
         \includegraphics[keepaspectratio,scale=0.5]{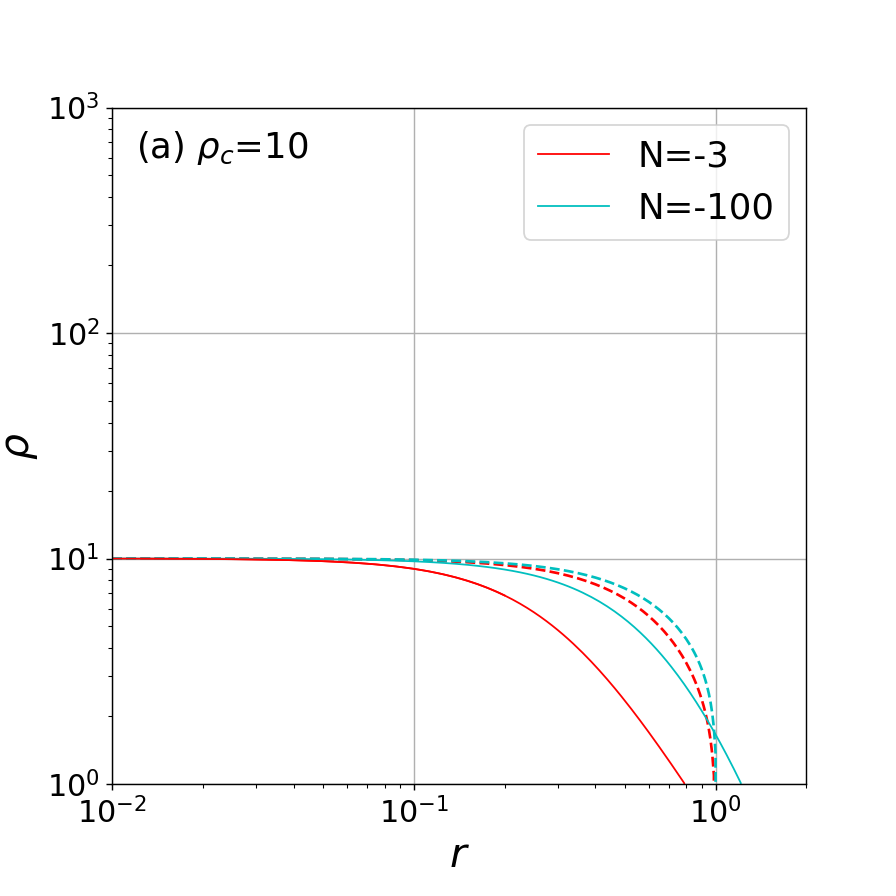} & \includegraphics[keepaspectratio,scale=0.5]{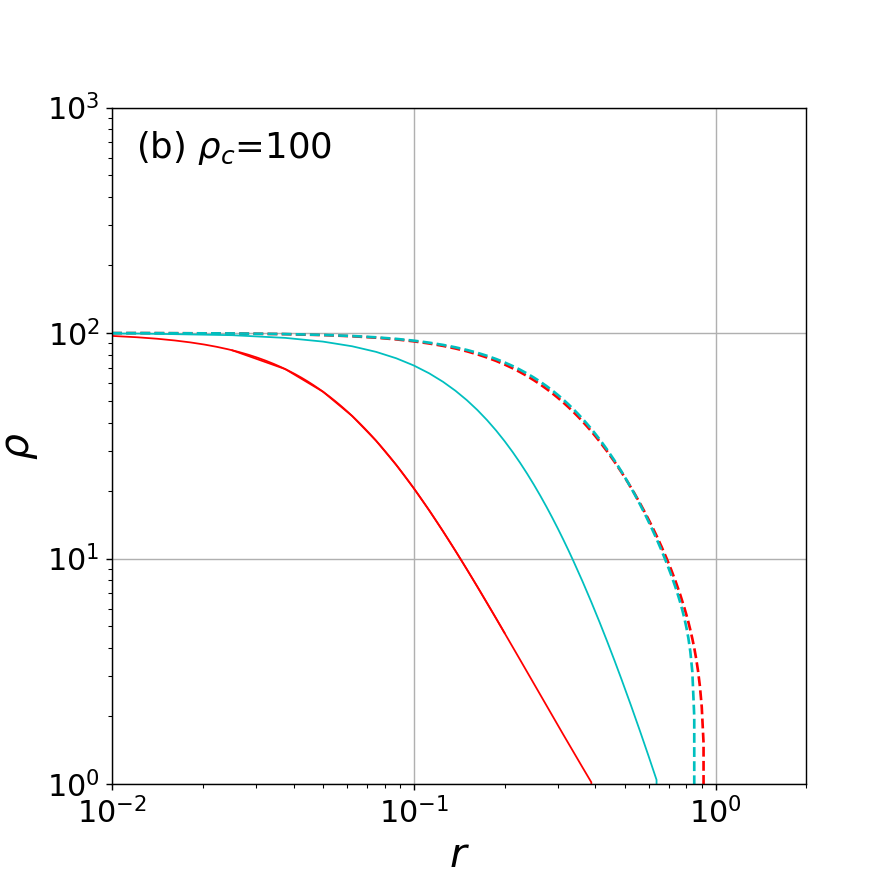}\\
         \includegraphics[keepaspectratio,scale=0.5]{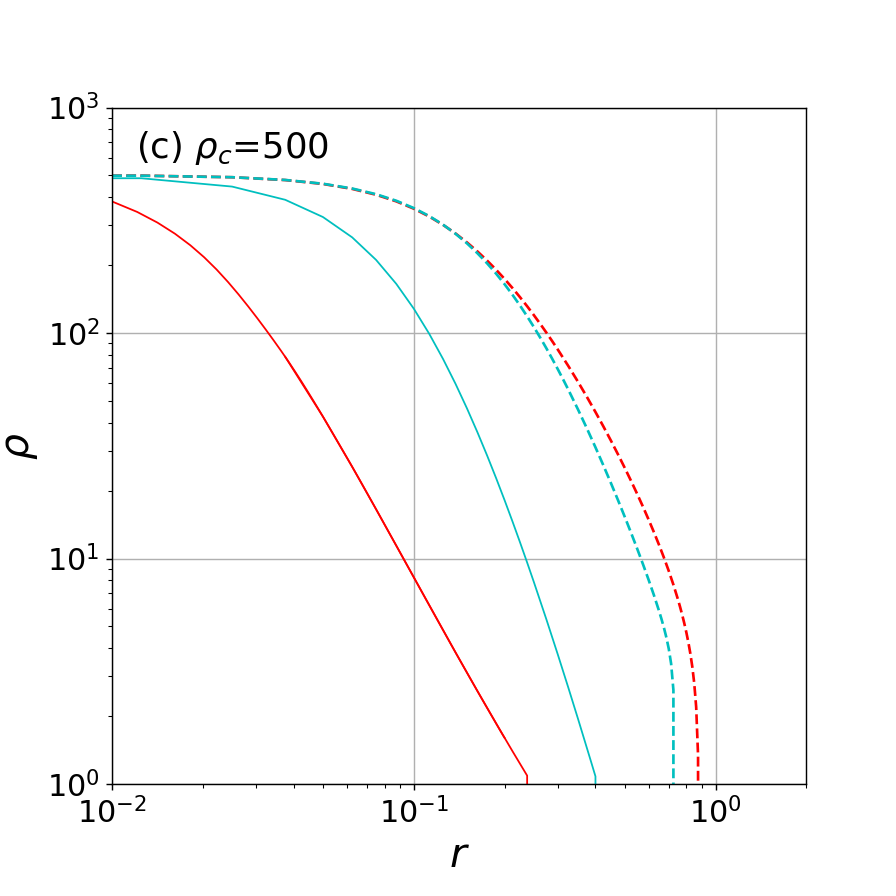}
     \end{tabular}
      \caption{ Comparison of the density profiles on the $x$- and $y$-axes for polytropic ($N=-3$) and isothermal ($N=-100$) filaments. Other parameters are constant: $R_0=1$ and $\beta_0=0.1$. The vertical axis shows the density, and the horizontal axis shows the distance from the center.  Red and cyan curves represent the models with $N=-3$ and $N=-100$, respectively. Solid and dashed curves correspond to the density profiles on the $y$- and $x$-axes, respectively. These panels correspond to different central densities:  $\rho_c=10$ (a),  $\rho_c=100$ (b), and  $\rho_c=500$ (c).\label{fig:density_R1}}
\end{figure}

Figure \ref{fig:density_R1} shows the density profiles on the $x$- and $y$-axes. 
In particular, the density profile on the $y$-axis clearly shows the effect of different $N$ values. 
Figure \ref{fig:density_R1} shows that the density distribution is divided into two parts: an inner core with an almost constant density and an outer envelope in which the density decreases with increasing distance from the center. This figure shows that both the isothermal and the polytropic filaments have power-law envelopes, except for the envelopes cutting along the $x$-axis for the $\rho_c=10$ models.
In terms of the distance to the surface from the center, the polytropic filament is more compact than the isothermal one in the $y$-direction.
In addition, the $\rho(y)$ distribution indicates that the density slope is shallower than the isothermal slope.
Although these two results seem to be inconsistent, this is natural if we consider that the compactness of a polytropic filament comes from the fact that it has a smaller core than an isothermal filament.

In contrast, density profiles on the $x$-axis are almost identical.
For example, the height of the polytropic filament on the $y$-axis is $40\%$ smaller than the isothermal height, while the $x$-axis width is $20\%$ wider than the isothermal width, when we compare density distributions with the same central density $\rho_c=500$. 

Next, we pay attention to the difference between the line-mass of each filament. For the central densities of $\rho_c=10,~100,~{\rm and}~500$, the line-mass of the polytropic and the isothermal filaments are obtained as  $\lambda_{N=-3}=10.36,~15.10,~{\rm and}~16.30$ and $\lambda_{N=-100}=17.90,~29.47,~{\rm and}~32.18$, respectively.
In both the polytropic and isothermal models, the line-mass increases as the central density increases. Meanwhile, comparison of filaments with the same central density shows that the polytropic filament is less massive than the isothermal one.
This is explained by the fact that the central pressure, which supports the filament against self-gravity, of the polytropic filament is smaller than that of the isothermal one.
For example, comparing two models with $\rho_c=100$, the central pressure of the polytropic model is only $p_c=100^{2/3}$, which is only $\sim21.5\%$ of the central pressure of the isothermal model.
Note that these properties come from the nature of the negative-indexed polytropic gas that is immersed in the same ambient gas pressure.

\subsection{Comparison of the Radius of the Parent Cloud}\label{sec:3.2}
In this section, we address the effect of the radius of the parent cloud $R_0$, which controls the magnetic flux threading the filament. The other parameters are constant at $N=-3$ and $\beta_0=0.1$.

\begin{figure}
\centering
     \begin{tabular}{lll}
         \includegraphics[keepaspectratio,scale=0.4]{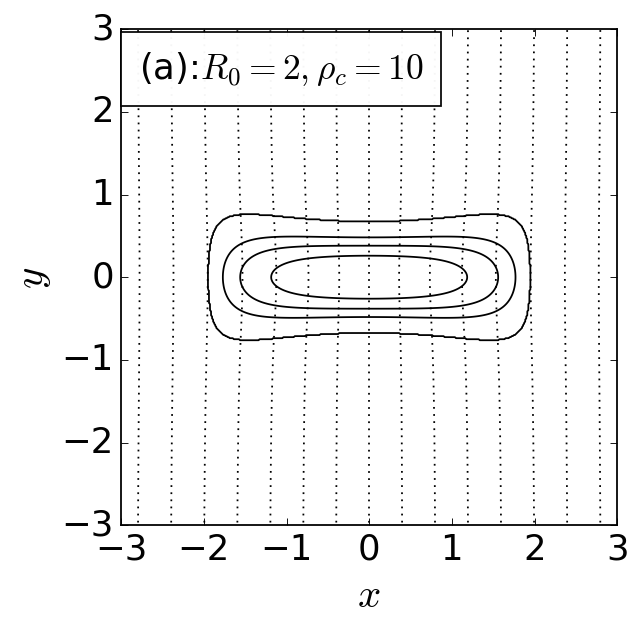}&
         \includegraphics[keepaspectratio,scale=0.4]{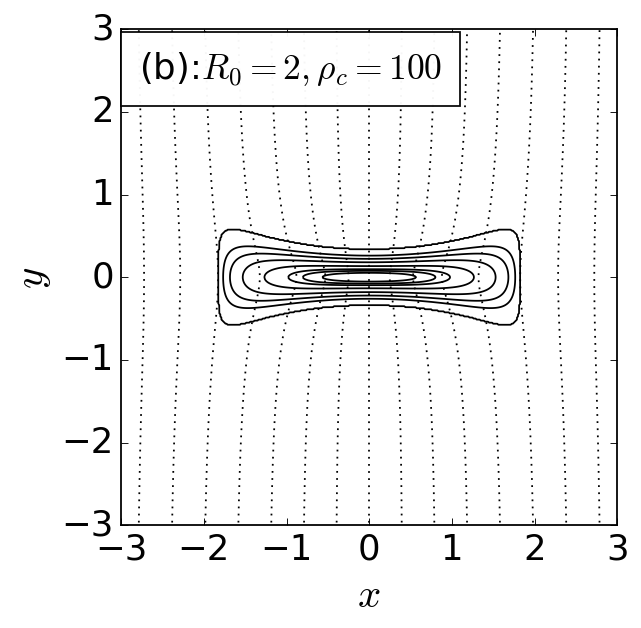}&
         \includegraphics[keepaspectratio,scale=0.4]{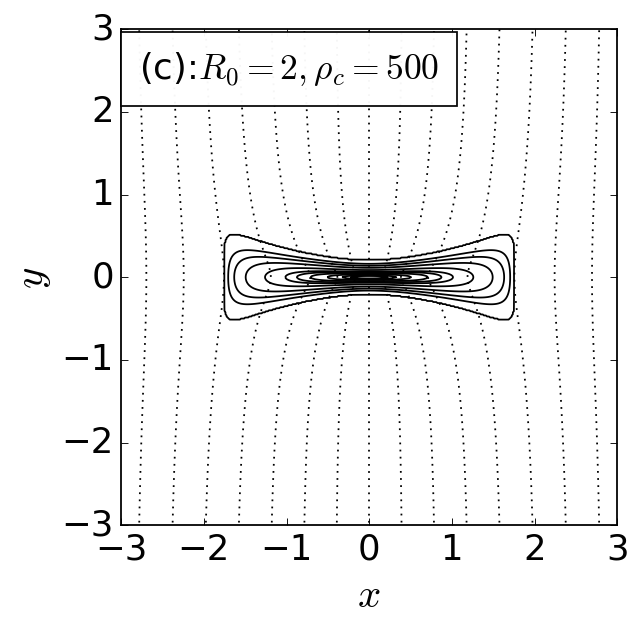}\\ 
         \includegraphics[keepaspectratio,scale=0.41]{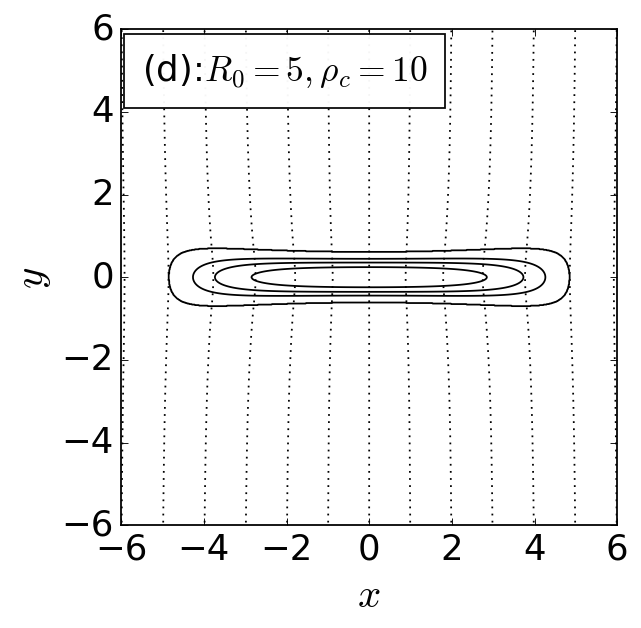}& 
         \includegraphics[keepaspectratio,scale=0.41]{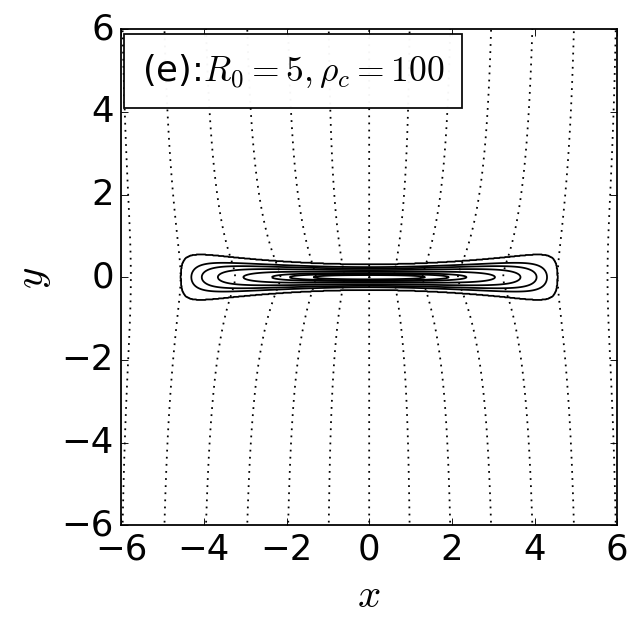}& 
         \includegraphics[keepaspectratio,scale=0.41]{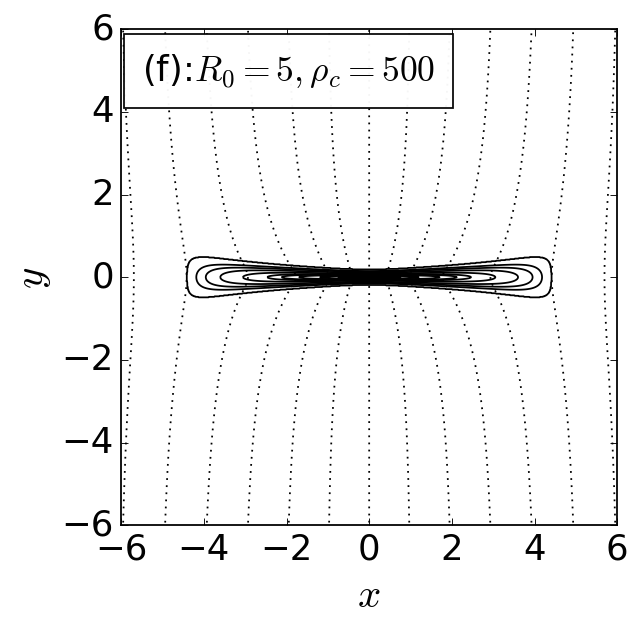}\\
     \end{tabular}
     \caption{Comparison of the models with different $R_0$ values: $R_0=2$ [(a)-(c)] and $R_0=5$ [(d)-(f)]. Other parameters are constant: $N=-3$ and $\beta_0=0.1$. The central density is equal to $\rho_c=10$ [(a) and (d)], $\rho_c=100$ [(b) and (e)], and $\rho_c=500$ [(c) and (f)]. The solid contour lines indicate the isodensity that is similar to that in Figure \ref{fig:cross_section_R1}. Dotted vertical lines are the magnetic field lines. The line-masses of these filaments are $\lambda_{R_0=2}=17.51~{\rm (a)},~26.68~{\rm (b)},~29.42~{\rm (c)}$ and $\lambda_{R_0=5}=39.31~{\rm (d)},~61.50~{\rm (e)},~68.61~{\rm (f)}$.
      \label{fig:cross_section_R2}}
\end{figure}

Figure $\ref{fig:cross_section_R2}$ shows the cross sections of the models of $R_0=2~\rm [(a) - (c)]$ and $R_0=5~ \rm [(d)-(f)]$ for respective central densities $\rho_c=10,~100,~{\rm and}~500$ (the model with $R_0=1$ is shown in the upper row of Fig.\,\ref{fig:cross_section_R1}).
When $\rho_c=500$, the height $y_s$ of the filament on the $y$-axis is equal to $y_s=0.238~(R_0=1),~y_s=0.213~(R_0=2), ~{\rm and}~y_s=0.188~(R_0=5)$ for the three different $R_0$ values, respectively.
In contrast, the half-width $x_s$ on the $x$-axis is equal to  $x_s=0.888~(R_0=1),~x_s=1.75~(R_0=2), ~{\rm and}~x_s=4.41~(R_0=5)$, respectively.
Thus the aspect ratio $x_s/y_s$ for $R_0=1$, $2$, and $5$ is equal to $x_s/y_s=3.73,~8.22,~{\rm and}~23.5$, respectively. Thus, the aspect ratio is an increasing function of $R_0$.

For the non-magnetic model of $\beta_0 =\infty$, we expected the cross section to be  round.
Nevertheless, the shape of the cross section is flat in the above models.
The magnetic field supports the filament in the $x$-direction but does not play a role in the $y$-direction. 
The average ratio of the Lorentz force to the thermal pressure force is equal to $3.27~(R_0=1),~ 6.25~(R_0=2), ~{\rm and}~15.3~(R_0=5)$ for $\rho_c=500$, respectively, measured on the $x$-axis. 
Thus, the Lorentz force is stronger than the thermal pressure, especially for the model with $R_0=5$. 
In addition, comparing three models with $\rho_c=10$, we found that the aspect ratio is $x_s/y_s=1.25~(R_0=1),~2.889~(R_0=2),~{\rm and}~8.083~(R_0=5)$ for the three different $R_0$ values, respectively. 
Models of $\rho_c=100$ indicate that the aspect ratio is $x_s/y_s=2.350~(R_0=1), ~5.407~(R_0=2),~{\rm and}~15.167~(R_0=5)$, respectively.
This shows that the aspect ratio increases as the central density increases when $R_0$ is the same.
Figure $\ref{fig:density_R2}$ shows the density profiles on the $x$- and $y$-axes.
The density profile on the $y$-axis is more compact than that on the $x$-axis, and the filament is flat. 
Comparison of the models with the same central density shows that the slope of the density profile on the $y$-axis is almost the same for the three different $R_0$ values. In contrast, the density profiles on the $x$-axis are not the same. The core radius on the $x$-axis increases as $R_0$ increases and, as a result, the distance to the surface also increases with increasing $R_0$. 

Next, we examine the difference in the line-mass. Figure $\ref{fig:lambda_rho_all_r}$ shows the relation between the line-mass and the central density for various $N$ and $R_0$ values with constant plasma beta $\beta_0=0.1$.
Comparison of models with the same central density and polytropic index show that the line-mass increases as $R_0$ increases.  
This suggests that the supported line-mass is controlled by the magnetic flux $\Phi_{\rm cl}=R_0\cdot B_0$.
This property is also valid for other polytropic indices.
Figure $\ref{fig:lambda_rho_all_r}$ also shows that the line-mass decreases with increasing polytropic index from $N=-100$ to $-3$, which is also discussed in section \ref{sec:3.1}.

\subsection{Effect of Plasma Beta}\label{sec:3.3}
Next, we compare the models with different $\beta_0$ values. Other parameters are fixed: $N=-3$, $R_0=2$, and $\rho_c=100$.

Figure \ref{fig:cross_section_beta} shows the cross section of the equilibrium state. Each panel corresponds to a different $\beta_0$ value: (a) $\beta_0=1$, (b) $\beta_0=0.5$, (c) $\beta_0=0.1$, and (d) $\beta_0=0.05$.
When $\beta_0$ is small and the magnetic field is strong, the magnetic field line retains its initial shape.
Because the filament is sufficiently supported by the strong magnetic field,
the surface of the filament is close to $R_0$ on the $x$-axis ($x_s\simeq R_0$). 

Figure \ref{fig:density_beta} shows the density profiles on the $x$- and $y$-axes for the same models shown in Figure \ref{fig:cross_section_beta}. 
The density profile on the $y$-axis is slightly affected by $\beta_0$.
In contrast, the density profile on the $x$-axis becomes steep in the models with low $\beta_0$. The strong Lorentz force extends the core radius but the width $x_s$ is not strongly affected by $\beta_0$.
Thus, the thickness of the envelope shrinks, which makes the slope steep.

Figure \ref{fig:lambda_rho_phi} shows the relation of the line-mass and the central density for filaments with $N=-3$ ($\triangle$) and $N=-100$ ($\circ$), in which different line colors represent different $\beta_0$ values.
All the models have the same $R_0=2$.
 Comparison of magnetized and non-magnetized polytropic filaments ($\triangle$ for $N=-3$) indicates that the line-mass of the magnetized filament 
 is heavier than the non-magnetized one (black symbols and solid curve).
Results for the isothermal filaments ($N=-100$) are the same. 
The line-mass of  the polytropic filament with $N=-3$ ($\triangle$) is smaller than that with $N=-100$  ($\circ$) when models with the same $\beta_0$ and $\rho_c$ are compared. This reflects the fact that the line-mass of the negative-indexed polytropic filament is less massive than that for the isothermal filament.
However, when the magnetic field is strong, the line-mass of the magnetized $N=-3$ filament is even larger than that of the non-magnetized isothermal filament ($N=-100$: grey symbols and solid curve).

\section{Discussion}\label{sec:discussion}
\subsection{Maximum Line-mass}
Section \ref{sec:3.1} shows that the line-mass decreases as $N$ increases from $-100$ to $-3$.
In section \ref{sec:3.2}, it is shown that the line-mass increases with increasing $R_0$.
The line-mass increases with decreasing $\beta_0$, as shown in section \ref{sec:3.3}.  
Thus, we expect the line-mass to be determined by the magnetic flux $\Phi_{\rm cl}\equiv R_0\cdot B_0$ and $N$. 
Here, we discuss how the maximum line-mass $\lambda_{\rm max}$ is expressed by the magnetic flux $\Phi_{\rm cl}$. 

The maximum line-mass $\lambda_{\rm max}$ represents the maximum allowable line-mass of a filament that is in equilibrium. When the line-mass $\lambda$ exceeds $\lambda_{\rm max}$, there is no equilibrium state and we call the filament 
``supercritical.'' In contrast, a filament with $\lambda<\lambda_{\rm max}$ is called ``subcritical,'' and its solution is discussed in the previous section \added{ (as a review, see \citet{2014prpl.conf...27A})}.

Although from its definition, $\lambda_{\rm max}$ is calculated as the slope $ \left(\partial \log \lambda / \partial \log \rho_c\right)_{N,R_0,\beta_0}=0$, considering numerical errors, we regard $\lambda_{\rm max}$ to be achieved when $\left(\partial \log \lambda/ \partial \log \rho_c\right)_{N,R_0,\beta_0} <0.05$ is satisfied.
\footnote{As indicated in Fig.~\ref{fig:lambda_rho_phi}, although in most of the models, $\lambda(\rho_c)$  monotonically increases with increasing $\rho_c$, some models [R2$\beta$1$(N=-3)$ and R2$\beta$0.5$(N=-3)$] have apparent peaks. These models with peaks enable  us to estimate the error in $\lambda_{\rm max}$ for a model in which $\lambda$ does not have a peak over the whole range of calculation, that is, $2\le \rho_c \le \rho_{c,\rm max}$. From model R2$\beta$1$(N=-3)$, the criteria $\left(\partial \log \lambda / \partial \log \rho_c\right)_{N,R_0,\beta_0}= 0.05$ and $=0.1$ yield line-masses approximately $1\%$ and a few percent smaller than the true maximum line-mass, respectively.}

Figure \ref{fig:maximumlinemass_phi} plots $\lambda_{\rm max}$ against $\Phi_{\rm cl}$ for various $N$ values.
This shows that $\lambda_{\rm max}$ increases with $\Phi_{\rm cl}$, and the slope seems almost the same at large $\Phi_{\rm cl}\agt 10$.
We approximated the curves by a function $\lambda_{\rm max}=\sqrt{\lambda^2_{0,\rm max}+A \Phi_{\rm cl}^2}$, where $\lambda_{0,\rm max}$ corresponds to the maximum line-mass of a non-magnetized polytropic filament and $A$ represents the slope, which is determined by fitting.
The value of $A$ for various $N$ values is obtained as $A=32.0~(N=-100),~26.4~(N=-10),~23.5~(N=-5),~{\rm and}~18.9~(N=-3)$.
From these four points,  
$A$ is fitted as $A=31.5+39.0/N$, and the accuracy of this fitting is within 5$\%$.
Thus, the normalized line-mass is expressed with $\lambda'_{0,\rm max}$ and $\Phi'_{\rm cl}$ as
\begin{equation}\label{eq:non_dim_empirical}
    \lambda'_{\rm max}\simeq \sqrt{{\lambda'}_{0,\rm max}^2+\left(31.5+\frac{39.0}{N}\right){\Phi'_{\rm cl}}^2},
\end{equation}
where $\lambda'_{0,\rm max}$ for various $N$ values is $\lambda'_{0,\rm max}(N=-100)=23.2$, $\lambda'_{0,\rm max}(N=-10)=17.5$, $\lambda'_{0,\rm max}(N=-5)=14.4$, and $\lambda'_{0,\rm max}(N=-3)=12.0$.
Because the line-mass and magnetic flux are normalized by $c^2_{\rm ss}/4\pi G$ and $c^2_{\rm ss}/(G/2)^{1/2}$, as in Table \ref{tab:normalization}, we obtain a dimensional form of Equation (\ref{eq:non_dim_empirical}) as
\begin{equation}\label{eq:dim_emp}
    \lambda_{\rm max} \simeq \sqrt{\left(\frac{\lambda'_{0,\rm max}(N)}{8\pi}\right)^2\left(\frac{2c^2_{\rm ss}}{G}\right)^2 +\left[\left(0.10+\frac{0.12}{N}\right)^{1/2}\frac{\Phi_{\rm cl}}{G^{1/2}}\right]^2}.
\end{equation}
Although the critical mass-to-flux ratio of the magnetized isothermal filamentary cloud is $(G^{1/2}\lambda/\Phi_{\rm cl})_{\rm crit}=0.24$ \citep{2014ApJ...785...24T}, the value for the negative-indexed polytropic filament ($N=-3$) is approximately equal to $0.24$.
Finally, Equation (\ref{eq:dim_emp}) is rewritten as
\begin{equation}\label{eq:empiri_last}
    \lambda_{\rm max} \simeq \sqrt{\left(\frac{\lambda_{0,\rm max}(N)}{M_{\odot} \rm pc^{-1}}\right)^2 +\left[5.9\left(1.0+\frac{1.2}{N}\right)^{1/2}\left(\frac{\Phi_{\rm cl}}{1\rm \mu G ~pc}\right)\right]^2}~~ M_\odot~\rm pc^{-1},
\end{equation}
where the maximum line-mass of a non-magnetized polytropic filament at each $N$ value is obtained as
\begin{equation}
       \lambda_{0,\rm max}(N)=
      \left\{
         \begin{array}{ll}
         15.5 \left(c_{\rm ss}/190\,{\rm m\,s^{-1}}\right)^2\,M_\odot\,{\rm pc^{-1}}&({\rm for\  }N=-100),\\
         11.7 \left(c_{\rm ss}/190\,{\rm m\,s^{-1}}\right)^2\,M_\odot\,{\rm pc^{-1}}&({\rm for\  }N=-10),\\
         9.6 \left(c_{\rm ss}/190\,{\rm m\,s^{-1}}\right)^2\,M_\odot\,{\rm pc^{-1}}&({\rm for\  }N=-5),\\
         8.0 \left(c_{\rm ss}/190\,{\rm m\,s^{-1}}\right)^2\,M_\odot\,{\rm pc^{-1}}&({\rm for\  }N=-3).
         \end{array}
         \right.
     \end{equation}
Thus, we derived an empirical formula of the maximum line-mass for magnetized polytropic filaments as Equation (\ref{eq:empiri_last}). For example, an $N=-3$ filament shows that the magnetic contribution for the line-mass becomes dominant when $\Phi_{\rm cl}>4.6\,{\rm pc\,\mu G}(c_{\rm ss}/190\,{\rm m\,s^{-1}})^2$.

\subsection{Column Density Distribution}\label{sec:column}
\begin{figure}
     \begin{tabular}{ll}
         \includegraphics[keepaspectratio,scale=0.5]{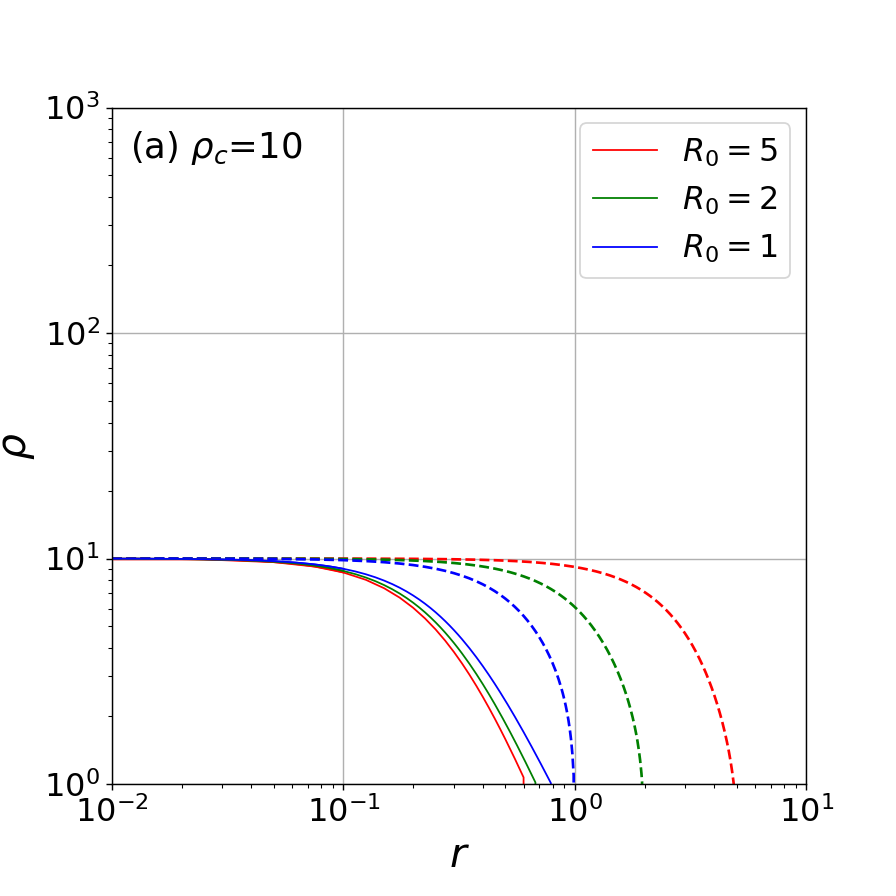} & \includegraphics[keepaspectratio,scale=0.5]{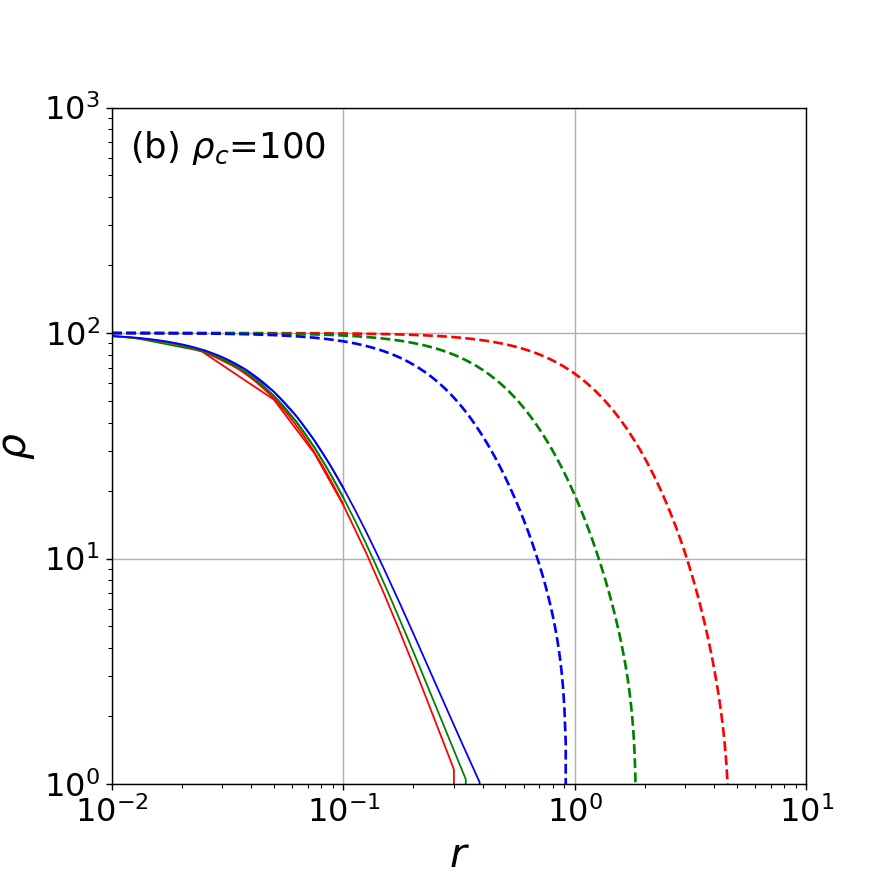}\\
         \includegraphics[keepaspectratio,scale=0.5]{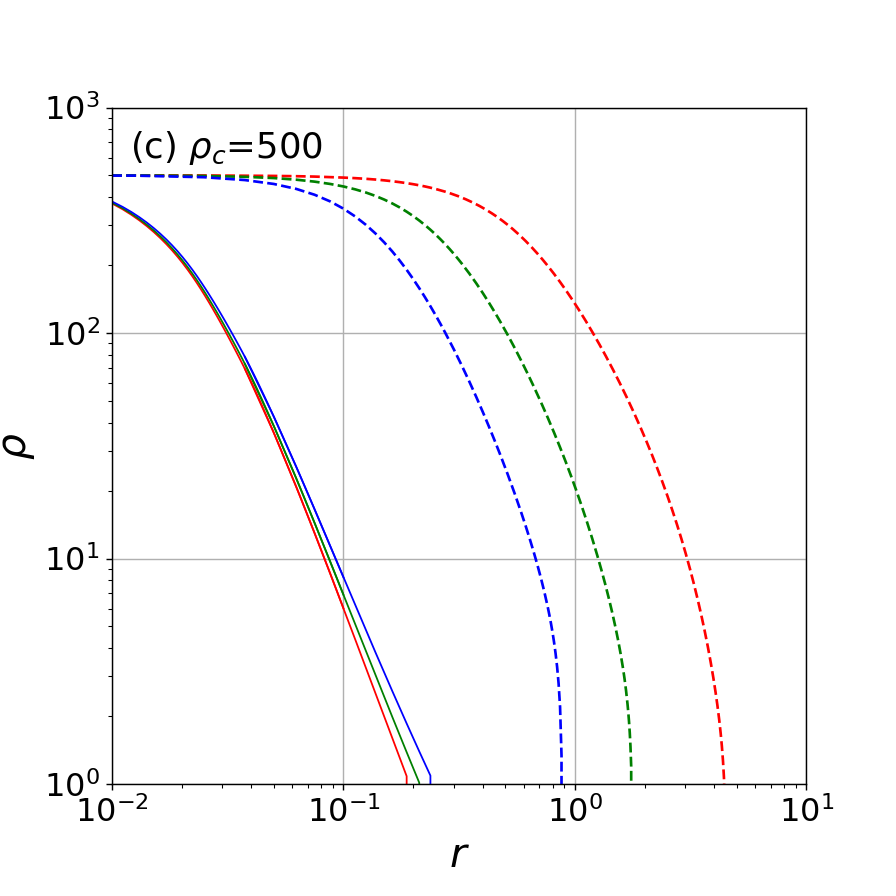}
     \end{tabular}
      \caption{Same as Fig. \ref{fig:density_R1} but for comparison of models with different $R_0$ values. Line color represents $R_0$ as $R_0=1$ (blue), $R_0=2$ (green), and $R_0=5$ (red).  Other parameters are constant: $N=-3$ and $\beta_0=0.1$. Panels (a), (b), and (c) correspond to the models with $\rho_c=10,~100,~{\rm and}~500$, respectively. Dashed and solid curves correspond to the density profile on the $x$- and $y$-axes, respectively.
      }
\label{fig:density_R2}
\end{figure}

\begin{figure}
    \centering
    \includegraphics[keepaspectratio,scale=0.7]{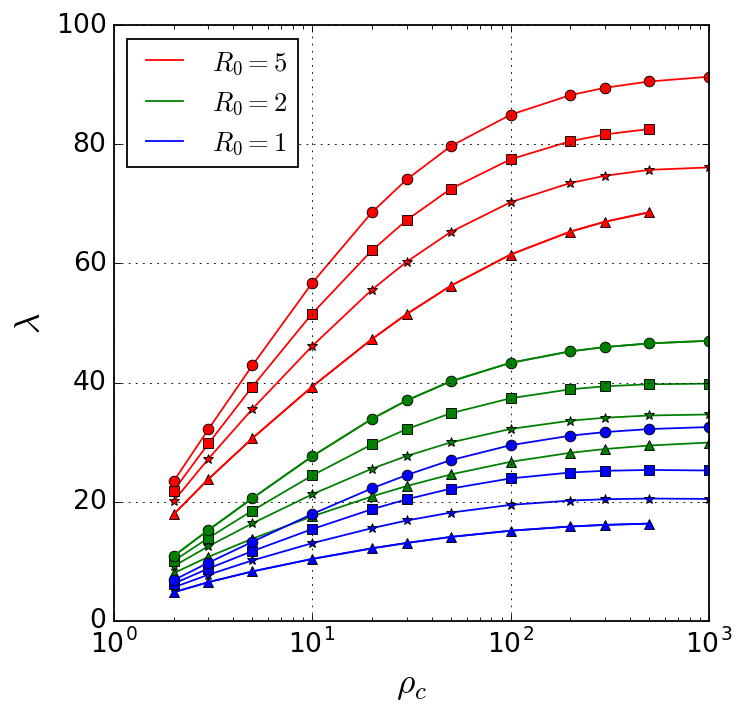}
      \caption{Line-mass plotted against the central density. Plasma beta is constant at $\beta_0=0.1$. The vertical and the horizontal axes represent the line-mass and the central density, respectively.  The colors blue, green,  and red correspond to the models with $R_0=1$, $2$, and $5$, respectively. The symbols $\circ$, $\sq$, $\star$, and $\triangle$ correspond to the models with $N=-100$, $-10$, $-5$, and $-3$, respectively. }
\label{fig:lambda_rho_all_r}
\end{figure}

\begin{figure}
   \begin{tabular}{cc}
         \centering
         \includegraphics[keepaspectratio,scale=0.6]{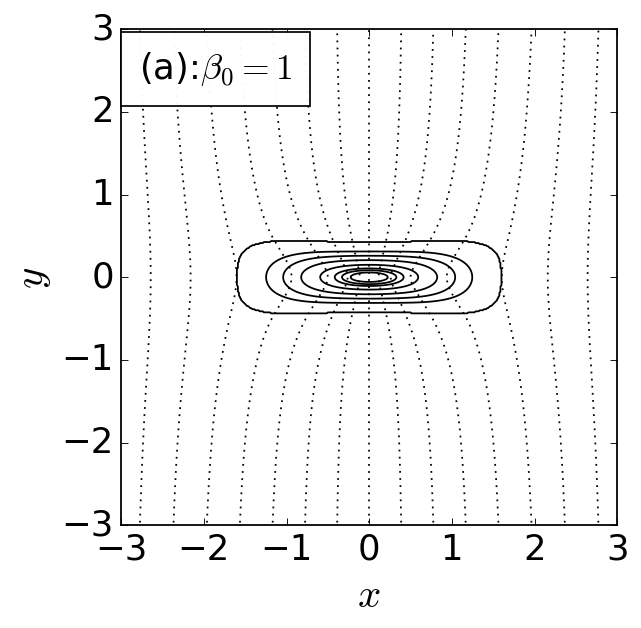} & 
        \includegraphics[keepaspectratio,scale=0.6]{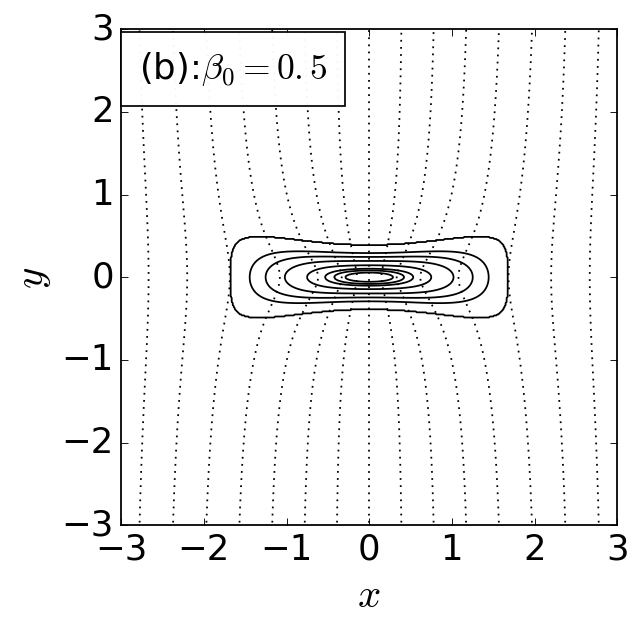}\\ 
       \includegraphics[keepaspectratio,scale=0.6]{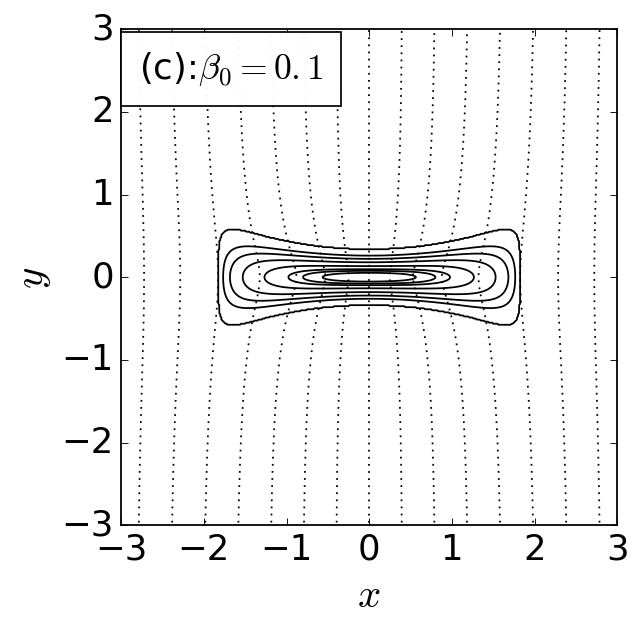}&
        \includegraphics[keepaspectratio,scale=0.6]{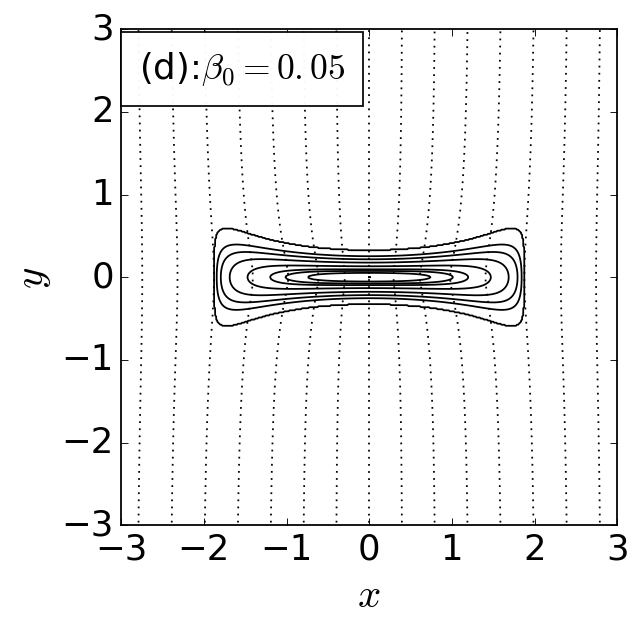}
        \\
     \end{tabular}
     \caption{Comparison of the solution models with different $\beta_0$ values. Other parameters are constant: $N=-3$, $R_0=2$, and $\rho_c=100$. Panels (a), (b), (c), and (d) correspond to $\beta_0=1$, $0.5$, $0.1$, and $0.05$, respectively. The line-masses of the respective models are $\lambda_{\beta_0=1}=14.26$ (a), $\lambda_{\beta_0=0.5}=17.22$ (b), $\lambda_{\beta_0=0.1}=26.68$ (c), and $\lambda_{\beta_0=0.05}=31.16$ (d).}
\label{fig:cross_section_beta}
\end{figure}

\begin{figure}
    \centering
    \includegraphics[keepaspectratio,scale=0.7]{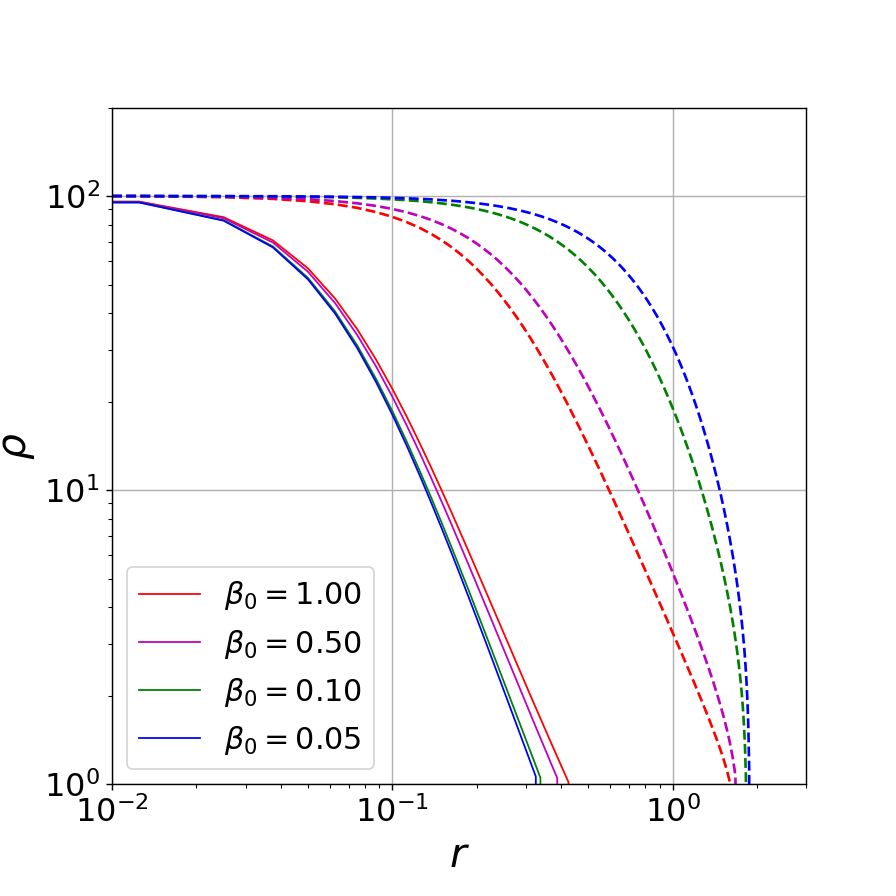}
      \caption{Density profiles on the $x$- and $y$-axes for the models shown in Figure \ref{fig:cross_section_beta}. Other parameters are constant: $N=-3$, $R_0=2$, and $\rho_c=100$. Line colors represent the models with different $\beta_0$ values as $\beta_0=1$ (red), $\beta_0=0.5$ (magenta), $\beta_0=0.1$ (green), and $\beta_0=0.05$ (blue). The solid and dashed curves show the density profiles along the $y$- and $x$-axes, respectively. }
\label{fig:density_beta}
\end{figure}

\begin{figure}
    \centering
    \includegraphics[keepaspectratio,scale=0.7]{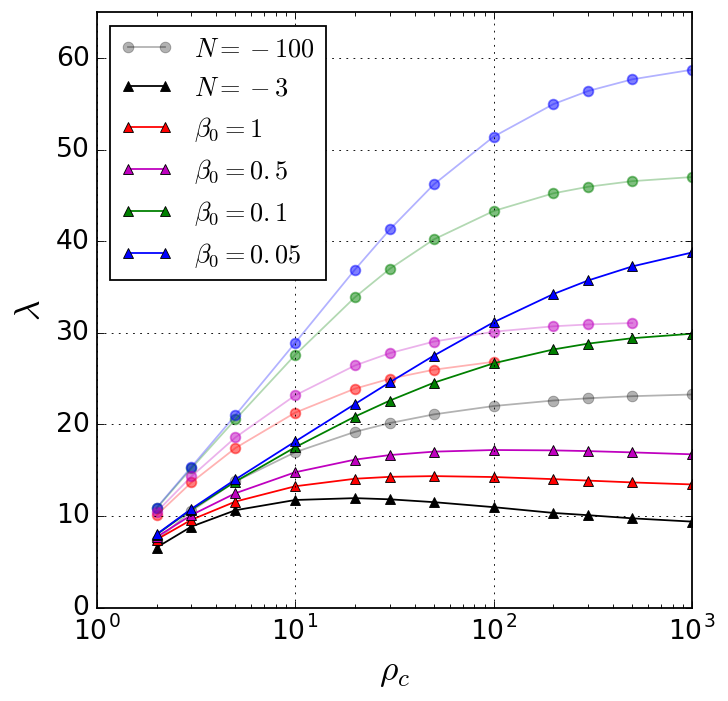}
      \caption{Relation between the line-mass and the central density for polytropic ($N=-3$) and isothermal ($N=-100$) filaments. Line colors represent models with different $\beta_0$  values as $\beta_0=1$ (red), $0.5$ (magenta), $0.1$ (green), and $0.05$ (blue). Thin colored curves with $\circ$ symbols and thick colored curves with $\triangle$ symbols represent, respectively, $N=-100$ and $N=-3$ models. The black ($N=-3$) and gray ($N=-100$) solid curves show results obtained from the non-magnetized Lane-Emden equation.
      }
\label{fig:lambda_rho_phi}
\end{figure}

\begin{figure}
    \centering
    \includegraphics[keepaspectratio,scale=0.7]{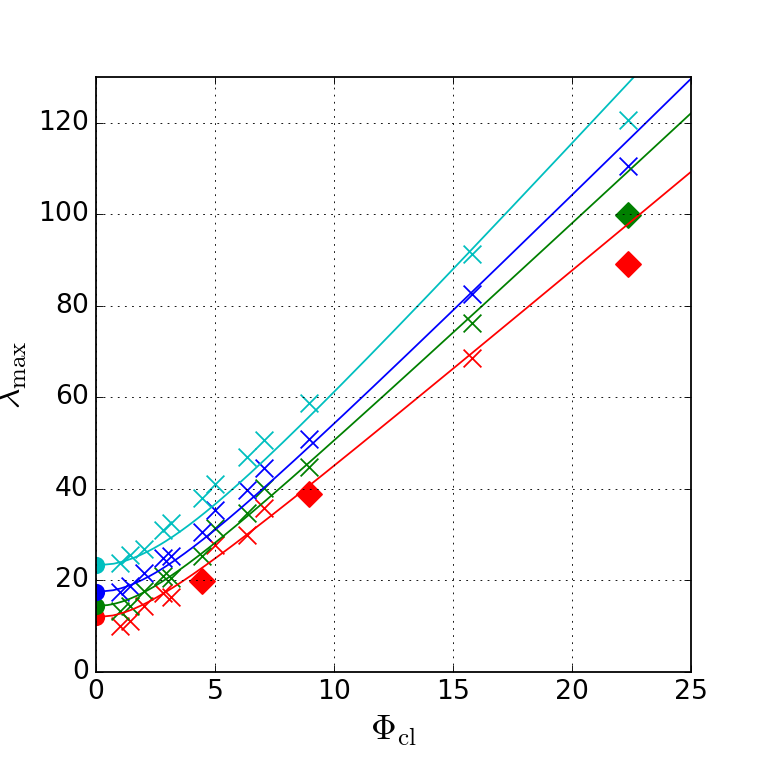}
    \caption{Maximum line-mass plotted against the magnetic flux.  The color represents different polytropic indices $N$ as cyan ($N=-100$), blue ($N=-10$), green ($N=-5$), and red ($N=-3$). The $\times$ symbols mean that the value is reliable, that is, obtained under the condition of $0\le\partial  \log \lambda/ \partial \log \rho_c\le 0.05$, while $\diamond$ symbols indicate that the value is the lower limit, $\partial  \log \lambda/ \partial \log \rho_c > 0.05$. Filled circles ($\bullet$) represent the maximum line-mass of the non-magnetized polytropic filament {$\Phi_{\rm cl}=0$}.}
\label{fig:maximumlinemass_phi}
\end{figure}

Figures $\ref{fig:density_R1}$, $\ref{fig:density_R2}$, and $\ref{fig:density_beta}$ show the density profiles of models with various $\rho_c$, $N$, $R_0$, and $\beta_0$ values.
The density profile on the $y$-axis is almost the same for three different $R_0$ and four different $\beta_0$ values. 
The slope of the profile on the $y$-axis becomes shallow when we increase $N$ from $-100$ to $-3$. 
Thus, the slope of the density profile is controlled only by the polytropic index $N$ in the $y$-direction, where the Lorentz force does not work.

This is understood as follows.
In the $y$-direction, because the Lorentz force does not play a role, the density distribution is governed by the pressure distribution (and self-gravity). 
A negative temperature gradient from the surface to the center has the effect of 
extending the envelope, and the temperature gradient increases from an isothermal equation of state ($N=-100$) to a polytropic one ($N=-3$).

Conversely, the slope on the $x$-axis becomes shallow only for the models with a weak magnetic field (Fig.~\ref{fig:density_beta}), while the slope slightly changes with $R_0$
(Fig.~\ref{fig:density_R2}).

We pay attention to these characteristics to consider how to reproduce the observed column density profile.
To characterize the density of the axisymmetric filament, Plummer-like profiles are often used, such as Equation (\ref{eq:plummer-like}).
Accordingly, the observed column density distribution is fitted with the function
    \begin{equation}\label{eq:plummer_fit}
        \sigma(r)=
        \frac{\sigma_0}{[1+(r/R_f)^2]^{(p-1)/2}},
    \end{equation}
    which is also a Plummer-like function,
    where $\sigma_0$, $R_f$, and $p$ ---
    the central column density, the core radius, and the density slope parameter, respectively --- are three fitting parameters \citep{2011A&A...529L...6A,2008MNRAS.384..755N}.
    We obtain the column density by integrating the numerical solution of the density distribution as
    \begin{subequations}
    \begin{align}
        &\sigma_\parallel (x)=2\int^{y_s(x)}_{y=0}\rho (x,y)dy,\\
        &\sigma_\perp (y)=2\int^{x_s(y)}_{x=0}\rho (x,y)dx,
    \end{align}
    \end{subequations}
where $\sigma_\parallel(x)$ and $\sigma_\perp(y)$ represent the column densities observed from the parallel and perpendicular directions, respectively, with respect to the magnetic field.
    
As shown in Equation (\ref{eq:density_ost}), the column density profile of an isothermal filament in hydrostatic equilibrium follows $p=4$ \citep{1963AcA....13...30S}. 
    However, as is summarized in section \ref{sec:intro}, $Herschel$ observations indicate that almost all the filaments follow $p\simeq 2$. 
    Although some researchers have argued that the cylindrical dynamical contraction explains the observed shallow column density slope of $p\simeq 2$ (such as \citet{1998PASJ...50..577K}), in the present paper, we investigate whether
    a hydrostatic filament having a negative temperature gradient
    forms the observed shallow column density slope.

\begin{figure}
     \begin{tabular}{ll}
     
         \includegraphics[keepaspectratio,scale=0.6]{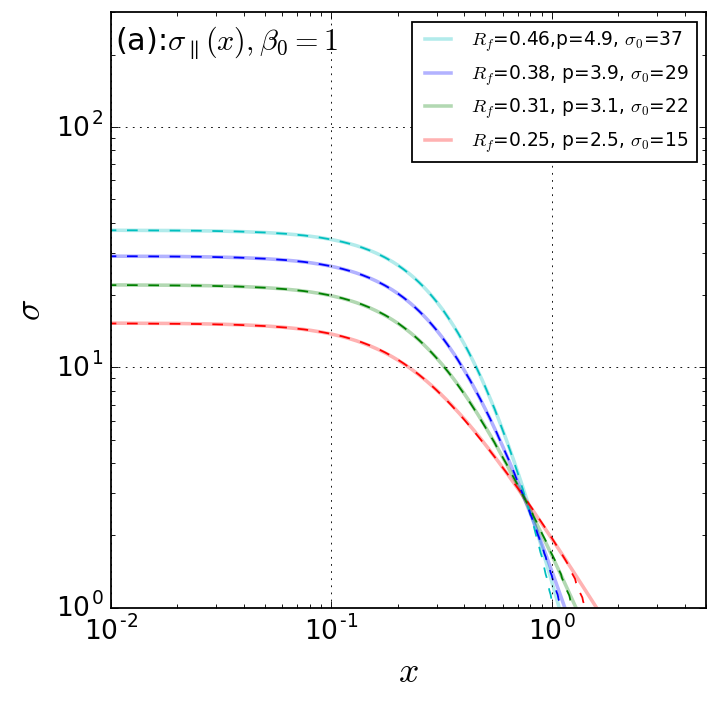} & 
         
        \includegraphics[keepaspectratio,scale=0.6]{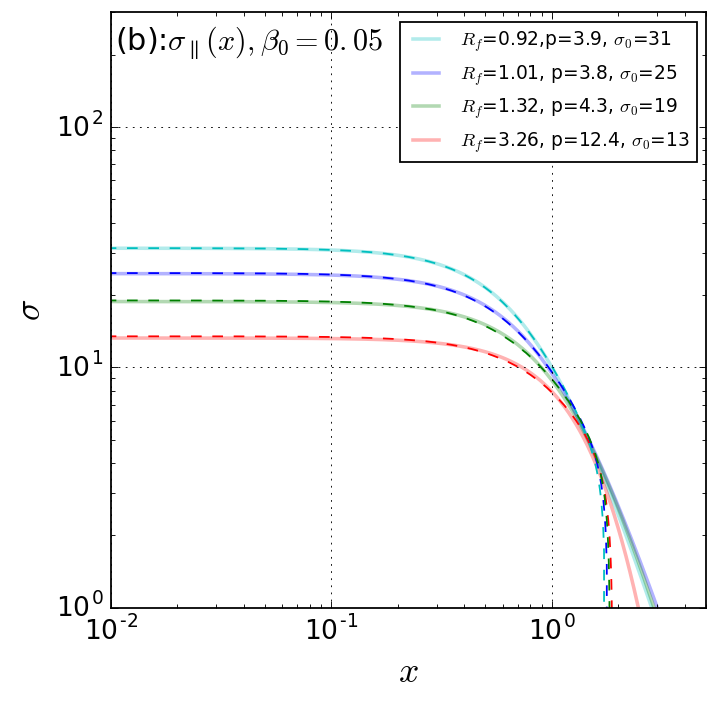}\\
        \includegraphics[keepaspectratio,scale=0.6]{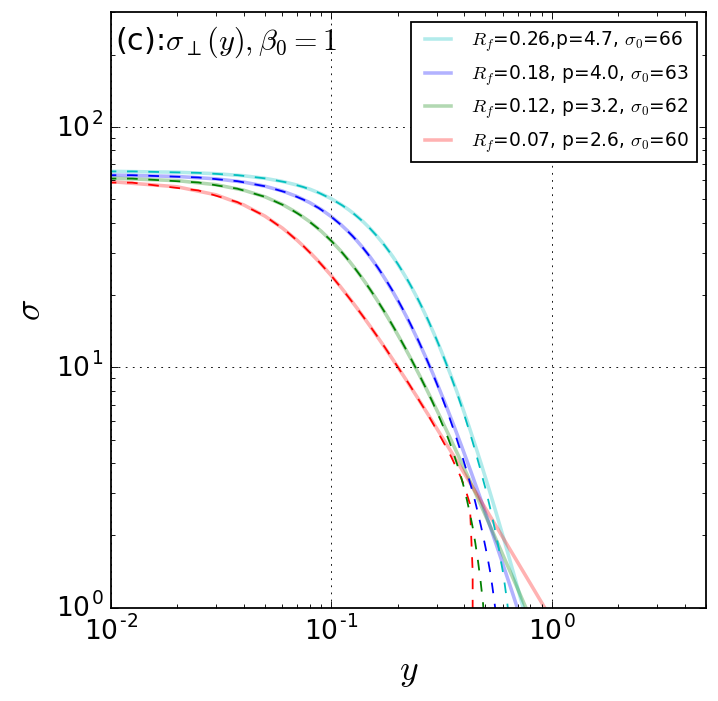} &
        
        \includegraphics[keepaspectratio,scale=0.6]{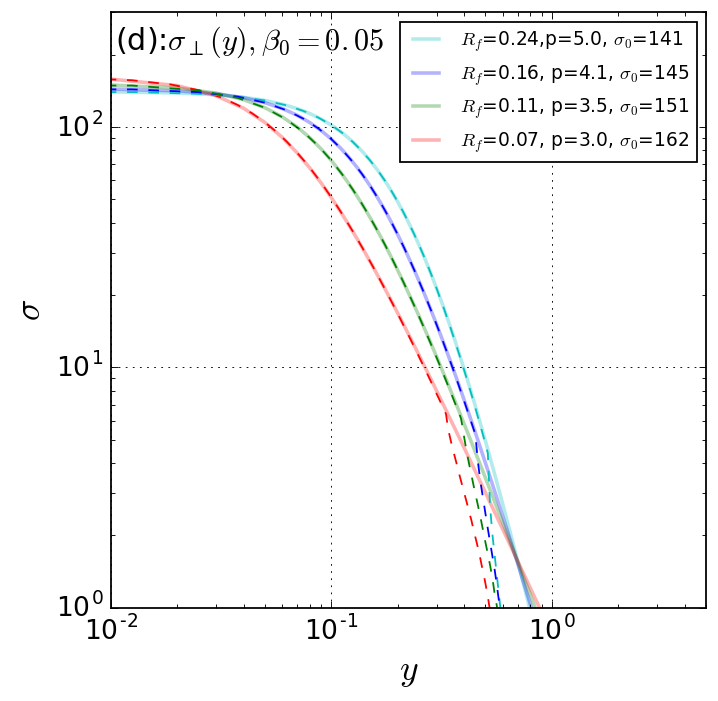}\\
     \end{tabular}
      \caption{ Fitted Plummer-like distributions: Each panel represents the column density distributions $\sigma_\parallel(x)$ [(a) and (b)] and $\sigma_\perp(y)$ [(c) and (d)] for various $N$ values. Panels (a) and (c) correspond to $\beta_0=1$, while (b) and (d) correspond to $\beta_0=0.05$.
      Other parameters are constant: $R_0=2$ and $\rho_c=100$.
      Dashed curves represent the column density distribution of magnetohydrostatic filaments, which are integrated along the $y$- and $x$-directions. Fitted Plummer-like distributions are shown by thin colored solid curves. 
      The color represents different polytropic indices $N$ as cyan ($N=-100$), blue ($N=-10$), green ($N=-5$), and red ($N=-3$).
      \label{fig:p_x} }
\end{figure}

    In Figure $\ref{fig:p_x}$, we show the column density integrated along the direction of the $y$- and $x$-axes
    for various $N$ and $\beta_0$ values. 
    Other parameters are constant at $R_0=2$ and $\rho_c=100$. 
    {We determined three parameters in the Plummer-like function [the slope index $p$, the column density at the center 
    $\sigma_0$, and the core radius $R_f$ in Eq.\,(\ref{eq:plummer_fit})] by the least squares method. 
    The least squares are calculated only for the region of 
    $\sigma \geq \sigma_0/10$. 
    This restriction comes from accounting for the dynamic range of the observed column density above the fore- and background column density  \citep{2019A&A...621A..42A}. }

    Figure \ref{fig:p_x} (a) and (b) presents the column density profiles of $\sigma_\parallel (x)$ that corresponds to 
    the profile in the direction in which the Lorentz force is effective. 
    When $\beta_0=1$, $p$ reaches 2 as $N$ goes from $-100$ to $-3$, $p(N=-100)=4.86 ~\rightarrow~p(N=-3)=2.48$ (panel a). Conversely, $p$ does not show such convergence when $\beta_0=0.05$. 
    For example, the model with $N=-3$ [red curve of Fig.~\ref{fig:p_x} (b)] indicates that the range of the power-law column density distribution is very narrow, and just outside of this, a sharp density decrement is observed.  If we try to fit this column density distribution with a Plummer-like function, this gives an artificially large power-law index $p$.
    
    In conclusion, the slope of the column density profile becomes shallow due to the temperature gradient for a model with a weak magnetic field.
    In contrast, we found that a strong magnetic field makes $R_f$ large and worsens the fitting with the Plummer-like function (\ref{eq:plummer_fit}). 
    
    Next, Figure \ref{fig:p_x} (c) and (d) corresponds to $\sigma_\perp(y)$, which indicates the column density distribution in the direction in which the Lorentz force does not play a role. 
    For both $\beta_0=1$ (panel c) or $\beta_0=0.05$ (panel d), the slope index $p$ reaches 2 as $N$ changes from $-100$ to $-3$.
    This resembles the relation obtained for the non-magnetized polytropic filament studied by \citet{2015MNRAS.446.2110T}.
    In the direction in which the Lorentz force is less important, the negative temperature gradient toward the center plays a role in making the density slope shallow, even in a magnetized filament.

\subsection{Column Density Distribution Depending on the Line of Sight}
\begin{figure}
    \centering
    \includegraphics[keepaspectratio,scale=0.7]{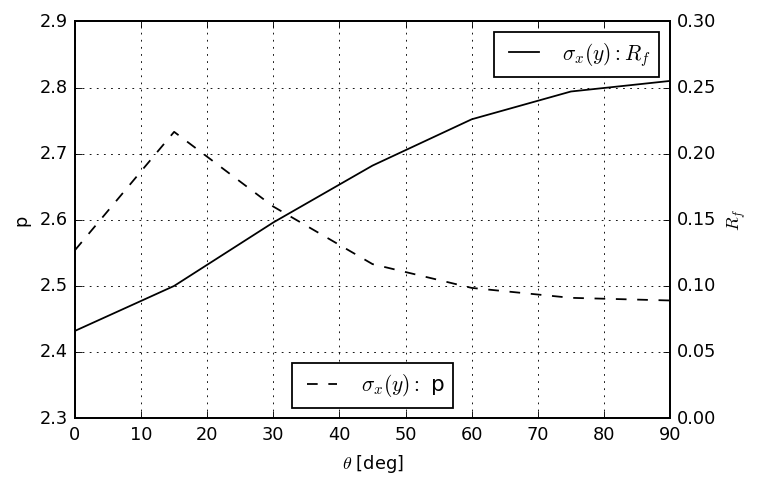}
      \caption{Core radius $R_f$ and density slope parameter $p$ plotted against $\theta$, which is the angle between the line of sight and the $x$-axis.
      The left and right vertical axes show $p$ and $R_f$, respectively.  
      The solid and dashed curves show the values of $R_f$ and $p$ at each $\theta$, respectively.}
\label{fig:rotation}
\end{figure}

In Figure \ref{fig:p_x}, we plot the column density distributions observed from the $y$-direction, $\sigma_\parallel(x)$, and that from the $x$-direction, $\sigma_\perp(y)$.
For comparison with observations, in this section we study the dependence of power-law index $p$ and core radius $R_f$ on the line of sight direction $\theta$, which is defined as the angle between the line of sight and the $x$-axis.  
Defining the angle between the line of sight and the $x$-axis as $\theta$,  we rotated the density distribution at $-\theta$. 
With the rotated density distribution integrated along the $x$-axis, $\sigma_x(y)\equiv\int \rho dx$ gives the column density distribution observed from this line of sight. 
In this case, the column density profiles $\sigma_\perp(y)$ and
$\sigma_\parallel(x)$ are obtained when $\theta=0^\circ$ and $\theta=90^\circ$, respectively.
Fitting the rotated $\sigma_x(y)$ with the Plummer-like function of Equation (\ref{eq:plummer_fit}), we obtain the power-law index $p$ and the core size $R_f$ depending on $\theta$,
as is shown in Figure \ref{fig:rotation}.
Parameters of the model are $R_0=2$, $N=-3$, $\beta_0=1$, and $\rho_c=100$.

Figure \ref{fig:rotation} shows that the power-law index $p$ (dashed curve) is restricted to a narrow range of $2.48\alt p\alt 2.73$.    
Thus, if we measure $p$, the line-of-sight direction is hardly determined. 
In other words, this seems to explain the reason why filaments commonly have $p\simeq 2$, even if the line of sight and thus the angle $\theta$ must be chosen randomly for each observational target. 
In contrast, the core size $R_f$ (solid curve) changes smoothly from 0.07 ($\theta=0^\circ$) to 0.26 ($\theta=90^\circ$).
Because $R_f$ is strongly dependent on $\theta$, we can distinguish whether the line of sight is nearly perpendicular ($\theta \simeq 0^\circ$) or
parallel ($\theta \simeq 90^\circ$) to the magnetic field.

\deleted{We now propose a way to distinguish whether $\theta=0^\circ$ or $\theta=90^\circ$. }
\added{We now propose a way to distinguish whether $\theta=0^\circ$ or $\theta=90^\circ$ when the line of sight is perpendicular to the filament long axis.}
In the non-magnetized and thus symmetric filament, Equation (\ref{eq:plummer_fit}) indicates that the central column density is equal to  $\sigma_0=S\rho_c R_f$, where the numerical factor $S=2\int_0^\infty (1+\zeta^2)^{-p/2}d\zeta$ equals $S=\pi$ for $p=2$ and $S=\pi/2$ for $p=4$.
This means that, in the axisymmetric model, the central column density is given as the central density times the scale length of the column density in the direction perpendicular to the line of sight.

In the non-axisymmetric configuration, we define the effective central density as
\begin{equation}
        \rho_{c}^{\rm eff} \equiv \frac{\sigma_0}{S \cdot R_f}.
\end{equation}

\begin{table}[htp]
\caption{Effective central density $\rho_c^{\rm eff}$, as calculated from the central column density $\sigma_0$ and core radius $R_f$ of the Plummer-like function fitting:  $\rho_c^{\rm eff}=\sigma_0/R_f/S$.  $\sigma_\parallel(x)$ and $\sigma_\perp(y)$ represent the column density distributions obtained by integrating the density in the directions parallel and perpendicular to the magnetic field, respectively.}
\begin{center}
\begin{tabular}{cccccccccc}
\hline\hline
  &$\beta_0$ & $N$&\multicolumn{3}{c}{$\sigma_\perp(y)$ }&&\multicolumn{3}{c}{$\sigma_\parallel(x)$}\\
  \cline{4-6}\cline{8-10}
  &&& $R_f$ & $\sigma_0$ & $\rho_c^{\rm eff}$  &&  $R_f$ & $\sigma_0$ & $\rho_c^{\rm eff}$\\ 
\hline
 &1& $-100$ & 0.26 & 55 & 67.4 && 0.46 & 37 & 25.6 \\
	&1 &$-10$ &0.18 &	63& 111 && 0.38 & 29 & 24.3 \\ 
	&1 &$-5$ & 0.12 &	62& 165 && 0.31 & 22 & 22.6 \\
	&1 &$-3$ & 0.07& 	60& 273 && 0.25 & 15 & 19.1 \\
\hline
 &0.05 &$-100$ & 0.24	& 141 &187   &&0.92&31 &10.7\\
&0.05 &$-10$&0.16&145&289		&&1.01	&25&7.88\\
&0.05 &$-5$&0.11&151&437		&&1.32	&19&4.58\\
&0.05 &$-3$&0.07&162&737		&&3.26	&13&1.27\\
\hline
\end{tabular}
\end{center}
\label{tab:tab3}
\end{table}%

Table \ref{tab:tab3} shows
the quantity calculated assuming $S=\pi$ for the models shown in Figure \ref{fig:p_x}.
Because all the models have the same central density $\rho_c=100$, $\rho_c^{\rm eff}$ derived from $\sigma_\parallel(x)$ is smaller than the true $\rho_c$.
For $N=-3$,  $\rho_c^{\rm eff}$ derived from $\sigma_\perp(y)$ is much larger than the true $\rho_c$,
but $\rho_c^{\rm eff}$ derived from $\sigma_\parallel(x)$
is much smaller than that.
When the central density is observationally estimated, for example,  by using the critical density of the molecular line transitions, we can compare this $\rho_c$ and $\rho_c^{\rm eff}$ estimated from
the central column density $\sigma_0$ and the column density scale length $R_f$.    
For a filament with $N=-3$,  $\rho_c^{\rm eff} \gg \rho_c$  for $\sigma_\perp(y)$.
Therefore, when  we observe $\rho_c^{\rm eff} \gg \rho_c$, this indicates that the line of sight is perpendicular to the magnetic field.
Conversely,  $\rho_c^{\rm eff} \ll \rho_c$  indicates that the line of sight is parallel to the magnetic field.
From this, when we observe the central density, central column density, and core radius of the filament, we can evaluate the angle between the line of sight and the magnetic field line.

\section{Summary and Conclusions}\label{sec:summary}
 We used the negative-indexed polytropic model to investigate the magnetohydrostatic equilibrium state of an interstellar filament with a lateral magnetic field and negative temperature gradient. Our findings are as follows:

\begin{enumerate}
\item Increasing the polytropic index from $N=-100$ to $-3$ flattens the filament cross section in a direction parallel to the magnetic field. 
When the density profiles of polytropic and isothermal filaments are compared in a direction parallel to the magnetic field, the envelope of polytropic filament is shallower. 
The line-mass of a polytropic filament is less massive in comparison to that of an isothermal filament when the filaments have the same central density and surface temperature.

\item When the radius of the parent cloud increases from $R_0=1$ to $5$, the aspect ratio of the cross-section (major-to-minor axis ratio) also increases.
Comparison of models with the same central density shows that the slope of the density profile parallel to the magnetic field is almost the same for three different $R_0$ values. In contrast, the density profiles perpendicular to the magnetic field are not the same because the core radius in that direction increases when $R_0$ increases. The line-mass increases with $R_0$ when we compare models with the same central density. 

\item Over the whole range of $\beta_0=0.05 - 1$, 
we found that the width of the filament in the direction perpendicular to the magnetic field is almost the same as that at $R_0$.
In this direction, a model with stronger magnetic field has a larger core radius than that of a weak magnetic model.
Thus, in such a model, the density profile in
the direction perpendicular to the magnetic field has a steep slope outside the core. Meanwhile, the density profile in the direction parallel to the magnetic field is almost the same irrespective of $\beta_0$. 
The line-mass becomes heavy with a small $\beta_0$ (strong magnetic field). 

\item We found that the maximum line-mass increases with the magnetic flux, and obtained the critical  mass-to-magnetic flux ratio as $\sim(0.10+0.12/N)^{1/2} G^{-1/2}$.
\item We conclude that a shallower column density profile is produced by a negative temperature profile in a magnetized filament. 
We succeeded in reproducing the observed column density profiles, especially in the direction where the Lorentz force is not effective or in the model with a weak magnetic field.
In the direction where the Lorentz force is effective, this mechanism does not work for a model with a strong magnetic field.

\item We proposed a way to estimate the angle between the line of sight and the magnetic field line.
We found that the core radius $R_f$ is strongly dependent on this angle.
This relation may help us distinguish whether the line of sight is nearly perpendicular or parallel to the magnetic field. 
\end{enumerate}

\acknowledgements{}  
The authors would like to thank Dr.\,K.\,Iwasaki for discussions on the model formulation and for careful reading of the manuscript.
This work was supported in part by a Grant-in-Aid for
Scientific Research (C) (No.\,19K03919) from the Japan
Society for the Promotion of Science (JSPS), in 2019-2021.

\bibliography{ref_polytropic_filament}{}
\bibliographystyle{aasjournal}   


%


\end{document}